\documentclass[sigconf]{acmart}

\usepackage{lipsum}
\usepackage{subcaption}

\usepackage{enumitem}
\setlist[itemize]{leftmargin=*}
\setlist[enumerate]{leftmargin=*,label=\arabic*.}

\usepackage[shortcuts]{extdash}

\usepackage{soul}

\usepackage{xspace}

\usepackage{xparse}

\AtBeginDocument{%
 \abovedisplayskip=3pt plus 3pt minus 3pt
 \abovedisplayshortskip=3pt plus 3pt minus 3pt
 \belowdisplayskip=3pt plus 3pt minus 3pt
 \belowdisplayshortskip=3pt plus 3pt minus 3pt
}

\newcommand{\code}[1]{{\small{\texttt{#1}}}}

\newenvironment{commentwrapper}[1]{\color{#1}}{\color{black}}

\definecolor{GREEN}{rgb}{0.0,0.7,0.0}
\definecolor{BLUE}{rgb}{0.0,0.2,0.7}
\definecolor{GOLD}{rgb}{0.6,0.6,0.0}
\definecolor{CYAN}{rgb}{0.0,0.6,0.6}
\definecolor{PURPLE}{rgb}{0.6,0.0,0.6}

\definecolor{RED}{rgb}{0.7,0.0,0.0}
\definecolor{GRAY}{gray}{0.5}

\definecolor{LIGHTGRAY}{HTML}{DDDDDD}

\usepackage{graphicx}
\usepackage{wrapfig}
\usepackage{makecell}
\usepackage{multirow}
\usepackage{algorithm2e}
\usepackage{csquotes}
\usepackage{colortbl}
\usepackage{fancybox}
\usepackage{array}
\usepackage{longtable}
\usepackage{booktabs}
\usepackage{pdflscape}

\definecolor[named]{ACMPink}{cmyk}{0.20, 0.60, 0, 0.20}
\definecolor[named]{ACMBrown}{rgb}{0.80, 0.46, 0.21}
\usepackage{acmart-taps}
\usepackage{pgfplots}
\AtBeginDocument{%
  \providecommand\BibTeX{{%
    \normalfont B\kern-0.5em{\scshape i\kern-0.25em b}\kern-0.8em\TeX}}}

\copyrightyear{2024}
\acmYear{2024}
\setcopyright{rightsretained}
\acmConference[CHI '24]{Proceedings of the CHI Conference on Human Factors in Computing Systems}{May 11--16, 2024}{Honolulu, HI, USA}
\acmBooktitle{Proceedings of the CHI Conference on Human Factors in Computing Systems (CHI '24), May 11--16, 2024, Honolulu, HI, USA}
\acmDOI{10.1145/3613904.3642400}
\acmISBN{979-8-4007-0330-0/24/05}

\newcommand{\sys}[0]{Luminate}

\begin{document}

\title[Prompting for Design Space]{Luminate: Structured Generation and Exploration of Design Space with Large Language Models for Human-AI Co-Creation}


\author{Sangho Suh}
\authornote{Contributed equally}
\orcid{0000-0003-4617-5116}
\affiliation{%
  \institution{University of California San Diego}
  \city{La Jolla}
  \state{CA}
  \country{USA}
}
\email{sanghosuh@ucsd.edu}

\author{Meng Chen}
\authornotemark[1]
\affiliation{%
  \institution{University of Notre Dame}
  \city{Notre Dame}
  \state{IN}
  \country{USA}
}
\email{mchen24@nd.edu}

\author{Bryan Min}
\orcid{0009-0003-0657-4398}
\affiliation{%
  \institution{University of California San Diego}
  \city{La Jolla}
  \state{CA}
  \country{USA}
}
\email{bdmin@ucsd.edu}

\author{Toby Jia-Jun Li}
\affiliation{%
  \institution{University of Notre Dame}
  \city{Notre Dame}
  \state{IN}
  \country{USA}
}
\email{toby.j.li@nd.edu}

\author{Haijun Xia}
\orcid{0000-0002-9425-0881}
\affiliation{%
  \institution{University of California San Diego}
  \city{La Jolla}
  \state{CA}
  \country{USA}
}
\email{haijunxia@ucsd.edu}






\renewcommand{\shortauthors}{Suh et al.}

\begin{teaserfigure}
  \includegraphics[trim=0cm 0cm 0cm 0cm, clip=true, width=\textwidth]{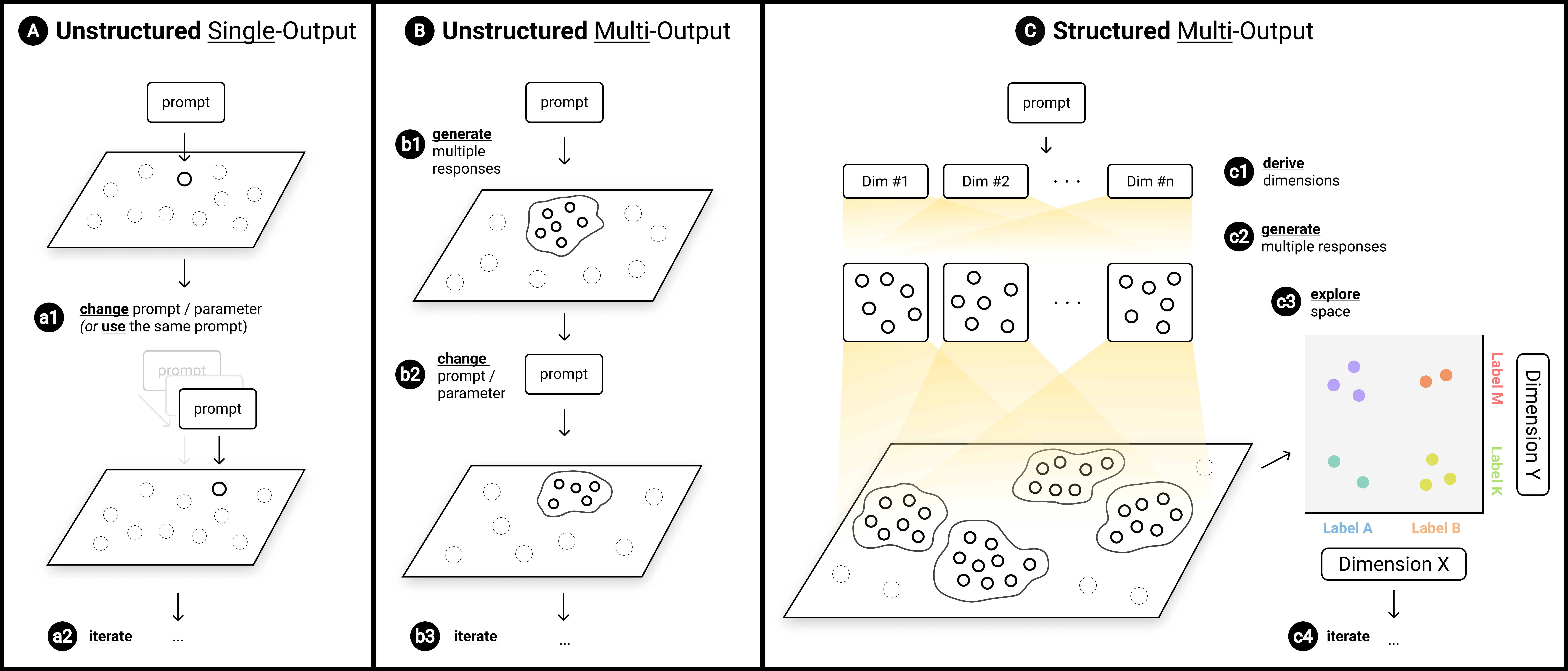}
    \caption{Our approach, structured multi-output (C), is shown with two current interaction paradigms (A \& B). We use \textit{structured} to denote the presence of dimensions relevant to the task / domain in guiding the response generation and \textit{unstructured} to denote their absence. Specifically, in our approach, users' prompt triggers (c1) generation of dimensions and subsequently the (c2) generation of responses using the dimensions from the previous step. Users can (c3) select dimension(s) to organize the responses in one- or two-dimensional space for exploration. The implementation and interaction details for each step are presented in the remainder of this paper.}
  \Description[short description]{long description.}
  \Description{Teaser figure}
  \label{fig:teaser}
\end{teaserfigure}

\begin{abstract}
Thanks to their generative capabilities, large language models (LLMs) have become an invaluable tool for creative processes. These models have the capacity to produce hundreds and thousands of visual and textual outputs, offering abundant inspiration for creative endeavors. But are we harnessing their full potential? We argue that current interaction paradigms fall short, guiding users towards rapid convergence on a limited set of ideas, rather than empowering them to explore the vast latent design space in generative models. To address this limitation, we propose a framework that facilitates the structured generation of design space in which users can seamlessly explore, evaluate, and synthesize a multitude of responses. We demonstrate the feasibility and usefulness of this framework through the design and development of an interactive system, Luminate, and a user study with 14 professional writers. Our work advances how we interact with LLMs for creative tasks, introducing a way to harness the creative potential of LLMs.
\end{abstract}



\maketitle

\section{Introduction}

A dictum from Linus Pauling, a two-time Nobel prize winner, states: ``\textit{if you want to have good ideas, you must have lots of ideas and learn to throw away the bad ones}'' \cite{kirwan2017s}. Indeed, creative processes typically begin with generating and exploring multiple ideas rather than refining a single one. Without first exploring diverse ideas, one can hastily converge to a \textit{sub-optimal} idea and spend time iterating on it without considering other options --- a phenomenon called fixation that hinders creative processes~\cite{jansson1991design, cross2004expertise, dow2010parallel}. Besides helping people avoid fixation, helping them understand the \textit{design space} --- the space of possible ideas and solutions to a task and problem --- has been found to play an essential role in creative processes~\cite{heape2007design, shaw2011role, dove2016argument, lomas2021design}. Thus, researchers have voiced the need to support \textit{design space thinking} in creativity support tools, stating various benefits of offering designers the ability to leverage an understanding of the design space in creative processes~\cite{schon1992designing, beaudouin2007prototyping, shaw2011role, biskjaer2014constraint, dove2016argument, halskov2021filtering}.

On the other hand, large language models (LLMs) today offer an unprecedented opportunity to empower creative processes. In addition to utilizing LLMs' generative power to automate and accelerate creation tasks~\cite{yuan2022wordcraft, kim2023metaphorian, gero2022sparks}, the capacity to instantly produce tens to hundreds of outputs is especially conducive to assisting users in developing a comprehensive understanding of the design space~\cite{halskov2021filtering, jonsson2022cracking}, a potential that current user interaction paradigms for LLMs (Fig.~\ref{fig:teaser}A \& B) are not fully harnessing. 
Instead, the current interaction paradigms help users \textit{converge} to an idea and refine the prompts (prompt engineering) to produce a more polished output from initial idea(s) they may have \textit{prematurely} converged to.

For example, recent text-to-image generation systems that generate multiple images for exploration focus on helping users refine their prompts to find the image that fits their desired quality or characteristics~\cite{brade2023promptify, feng2023promptmagician}. 
Specifically, they do not provide a design space for reflection and exploration to facilitate and encourage \textit{design space thinking}. Instead, after users add their prompt (e.g., subject), they are quickly presented with a set of similar results (e.g., images about the subject), confining them within a narrow design space. 
Although this approach, illustrated in Fig.~\ref{fig:teaser}B, allows users to explore alternative ideas, the generation of these outputs is \textit{unstructured} (i.e., the method does not employ a systematic approach, such as structuring the generation of responses using key dimensions and values that make up the design space) and the exploration of these outputs encourages users to converge rather than diverge~\cite{dow2010parallel} --- an argument we further elaborate on in Section~\ref{sec:framework_motivation}.

The goal of this work, therefore, is to explore a new interaction paradigm with LLMs that scaffolds and encourages design space thinking in the human-AI co-creation process. To achieve this, we draw insights from prior work in exploratory search~\cite{ahlberg1994visual, marchionini2006exploratory, schraefel2006mspace}, creativity support~\cite{jansson1991design, buxton2010sketching, dow2010parallel}, dimensional reasoning~\cite{dove2016argument, macneil2017dimensional}, and information visualizations~\cite{villa2009aspectual, endert2012semantic}. These fields repeatedly demonstrated that explicit enumeration of, visualizing data within, and interaction with the principle dimensions associated with the target domains or tasks can enhance understanding and exploration of the design space. Based on these insights, we propose a human-AI collaboration framework (Fig.~\ref{fig:teaser}C). Instead of generating responses directly based on the user's prompt, we first prompt LLMs to generate key dimensions associated with the topic or task in the original prompt, and then combine the derived dimensions with the original prompt to systematically generate responses to construct the design space. Users can then leverage these derived dimensions to explore the generated design space in a structured, systematic manner.

To demonstrate the feasibility and usefulness of this framework, we instantiated it by developing \sys{}, an interactive system that uses LLMs to first generate key dimensions from a user's prompt and then generates responses using these dimensions for structured exploration of LLM outputs. Our user study with 14 professional writers specializing in creative writing shows that our approach has the potential to help users effectively explore the design space during their creative processes. Our work contributes a step towards improving the interaction and usefulness of LLMs in assisting creative processes. In summary, our contributions include:

\begin{itemize}
    \item A new interaction framework for human-AI collaboration in creative tasks that build on the premise that LLMs should allow users to explore a space of possible responses, rather than giving a single data point in response to user input. 
    \item \sys{},\footnote{\url{https://luminate-research.github.io/}} an interactive system that demonstrates this idea with novel interaction techniques and features for structured generation and exploration of LLM outputs.
    \item A user study demonstrating that enabling dimensional exploration of LLM output space has the potential to prevent fixation and nurture design space thinking.
\end{itemize}
\section{Background}
\label{sec:background}

We present prior research and concepts relevant to our work, namely supporting creative process, co-creating with AI, and facilitating visuo-spatial exploration and sensemaking.

\subsection{Supporting Creative Process}
\label{sec:supporting-creative-process}

\textit{Divergent-convergent thinking.} Creative thinking entails two modes of thought: divergent and convergent thinking~\cite{guilford1961three, willemsen2023role}. Divergent (\textit{lateral}) thinking is the process of generating multiple, unique ideas or solutions to a problem or task, whereas convergent (\textit{vertical}) thinking is the process of integrating, synthesizing, and evaluating ideas and possibilities to arrive at (dive into) a single (creative) outcome~\cite{de1970lateral}.
Because divergent thinking is responsible for the generation of ideas and convergent thinking the evaluation of these ideas, prior research established that the interplay between the two is imperative in creative processes and noted that creativity support tools should support the constant cycle of these two modes in creative processes~\cite{guilford1967nature, brophy2001comparing}. Despite the benefits of divergent thinking, such as avoiding fixation, leading to better and more diverse outcomes, and improved self-efficacy \cite{dow2010parallel}, studies found that even experts often converge prematurely during creative processes~\cite{jansson1991design, cross2004expertise}, suggesting the human tendency to converge and challenges to diverge.

\textit{Design space.} 
With divergent-convergent thinking, design space plays a critical role in creative processes~\cite{westerlund2005design, beaudouin2007prototyping, biskjaer2014constraint}. The term design space is generally used to refer to a conceptual `space of possibilities' --- broadly construed as collections of ideas, designs, concepts, or solutions~\cite{maclean1991design, halskov2021filtering}. Design spaces have commonly been represented with Cartesian spaces where possible variations reside within the dimensions of a design, and variations reflect different values on those dimensions~\cite{beaudouin2007prototyping, shaw2011role, halskov2021filtering, lomas2021design}. For example, the design space of a narrative might include dimensions such as \textsf{genre}, \textsf{setting}, \textsf{point of view}, \textsf{tone}, and so on. Each narrative in the space would reflect the values in those dimensions (e.g., \textit{fantasy} for \textsf{genre}, \textit{modern times} for \textsf{setting}, etc.) and positioned accordingly in the Cartesian space. 

\textit{Dimensional Reasoning.} Since design space makes dimensions and their values explicit, it enables designers to reflect on the problem space, desired direction and characteristics, and to systematically evaluate different possibilities~\cite{beaudouin2007prototyping}. Essentially, it enables dimensional reasoning, allowing people to step back from the individual idea, artifact, or solution and view the entire problem space at a conceptual, higher level. This offers an all-encompassing view~\cite{heape2007design}, which is why design space is frequently used in HCI and other design-related disciplines to classify new artifacts, as it makes it easy to compare them with existing artifacts or designs to understand what makes them different or novel (e.g.,~\cite{wiberg2014makes}). 

Simultaneously, design space plays a crucial role in generating new designs, ideas, and artifacts~\cite{macneil2017dimensional, generativetheoryofinteraction}. 
While a design space ``constrains design possibilities along some dimensions, it leaves others open for creative exploration''~\cite{beaudouin2007prototyping}. This can be seen, for example, in research that uses design cards to generate new designs~\cite{golembewski2010ideation, mora2017tiles, suh2020coding, suh2022privacytoon}. 
For instance, Lomas et al. created design space cards for game designs where elements of game design theory (e.g., Action, Theme, Play Styles) were used as dimensions. They found that these design space cards can help a range of audiences --- from children and design novices to experts --- to effectively create new game designs~\cite{lomas2021design}. Other similar research that created design cards to support design in their respective areas reported the same findings~\cite{mora2017tiles, bach2018design, suh2020coding, suh2022privacytoon}.

As such, while thinking with dimensions offers many benefits, prior work found that dimensional reasoning does not happen naturally and that even experts can have difficulties answering the dimensions that characterize the problem space they work in~\cite{macneil2017dimensional}. 
This suggests that in order to encourage and enable dimensional reasoning, we need to explicitly surface dimensions to users' attention and support the generation of these dimensions.

Our work builds on these lines of work and extends it in several ways. First, our framework demonstrates how we can interweave the idea of design space (thinking) into the LLM-powered workflow to enhance the usefulness of LLMs in creative processes. Second, we show the use of LLMs to generate dimensions for dimensional reasoning in creative processes, which should bring the benefits of dimensional reasoning while solving the challenge users have with identifying or recalling key dimensions characterizing the problem space, task, and domain. (We showcase these interactions through our system in Section~\ref{sec:system}.)

\subsection{Co-Creating with AI}

Co-creation with AI is rapidly gaining traction as a prevalent approach in creative processes. This collaborative style of work has already found application in numerous domains, such as music composition~\cite{louie2020novice}, creative coding~\cite{jonsson2022cracking}, writing~\cite{yuan2022wordcraft, gero2022sparks, kim2023metaphorian, zhang2023visar, yang2022ai}, design~\cite{liu2022opal, jiang2022promptmaker, lawton2023drawing, lu2022bridging}, and video authoring~\cite{wang2023reelframer}. Indeed, much work focused on demonstrating how to co-create in these particular domains, as well as understanding the challenges in co-creating with AI~\cite{gmeiner2023exploring} and enabling users to more effectively control and steer AI's output (e.g., prompt engineering~\cite{wu2021ai, kim2023metaphorian, brade2023promptify}).

Amidst the growing interest in this partnership, there has been a long-lasting theme of research to explore how AI can be effectively intertwined with humans so that AI augments --- rather than fully automates --- our productivity and creativity~\cite{licklider1960man, horvitz1999principles, amershi2019guidelines}. Licklider proposed human-AI symbiosis, where the human defines the goals and criteria, and AI prepares the tasks for the human to perform more insightful work \cite{licklider1960man}. Furthermore, to build a comprehensive perspective on human-AI co-creation, efforts have been made to chart the design space in the context of text generation tasks~\cite{cheng2022mapping, ding2023mapping}. Cheng et al. conducted a systematic analysis of the design space, culminating in the creation of a taxonomy featuring five dimensions, including human actions, types of human control, model iterations, workflow initiation, and design of interfaces between human and AI ~\cite{cheng2022mapping}. Building on this taxonomy, Ding and Chan identified a spectrum of human-AI co-creation tasks, each associated with specific patterns of human-AI interaction~\cite{ding2023mapping}. Within this spectrum, they placed human-AI co-creation tasks according to the degree of required human intervention, from minimal to substantial. They concluded that complex and interdependent creative tasks, such as paper and fiction writing, thrive in human-AI co-creation due to the iterative and interdependent nature of the interactions.

Our work builds on the extensive exploration of human-AI collaboration research, where they leverage the understanding of the unique strengths and weaknesses of humans and AIs. Our perspective and goal are different from these prior research in that, instead of focusing on leveraging human-AI collaboration to accomplish a creation task, our primary goal is to leverage AI to support humans in developing a more comprehensive understanding of the design space so that the human can more effectively identify the requirements, criteria, and constraints of the creativity task, which was often assumed to be the responsibility and difficulty that the human need to take on their own \cite{licklider1960man, macneil2017dimensional}.

\subsection{Facilitating Exploration and Sensemaking}
While LLMs' ability to generate a large number of responses presents an opportunity for creativity support, it also poses a challenge in how to visualize the responses without overwhelming users and help them navigate the design space. To address this challenge, we build upon visualization and interaction techniques that have been explored in information visualization and HCI. 

Visualization techniques for multivariate data visualization --- such as interactive scatter plots, linked axes \cite{claessen2011flexible}, and small multiples \cite{maceachren2003exploring, bezerianos2010graphdice} --- enable users to inspect the distribution of data points across individual or pairs of dimensions. For example, with linked axes, each dimension of the data set forms an axis, and data points are visualized as links that connect the various axes \cite{claessen2011flexible}. Small multiples with scatter plots display a series of scatter plots showing the data distribution across pairs of data dimensions \cite{bezerianos2010graphdice}. Such systems represent vast amounts of data in a compact 2-dimensional format and empower users with interactive features such as searching, filtering, and user-driven clustering of data points to enable flexible exploration and facilitate the understanding of the data set. 

While representing rich multivariate data points as dots on a scatter plot enables users to easily inspect the distribution of data points, such abstraction also makes it challenging to dive into the details of specific data points. Semantic zooming is a technique that has been used to address this. It suggests visualizing different levels of details at various levels of zooming scales \cite{bederson1994pad++}. This technique enables users to flexibly switch between the overview of the space to the detailed view of specific data points, facilitating the understanding of focused information in context \cite{card1999readings}, and has been used for a variety of sensemaking tasks \cite{suh2023sensecape, muthukumarasamy1995visualizing, hayatpur2023crosscode}. Closely related to our context is Sensecape, which leverages semantic zoom to enable users to easily overview large amounts of text responses generated by LLMs, where text responses flexibly transition across three different levels of abstraction --- keywords, summary sentences, and full text --- at different zooming scales~\cite{suh2023sensecape}.

Our work builds on these well-known visualization and interaction techniques formulated to facilitate the exploration and sensemaking for the large dataset of multivariate responses generated by LLMs. We leverage the interaction visualizations as graphical interfaces to enable users to request more responses with existing dimensions or re-evaluate existing responses with new dimensions to facilitate the iterative divergent and convergent creative processes --- novel interaction techniques shown with \sys{} (Section~\ref{sec:system}) that we believe can contribute to advancing Human-AI co-creation.

\begin{figure*}[htb!]
    \centering
    \includegraphics[trim=0cm 0cm 0cm 0cm, clip=true, width=\textwidth]{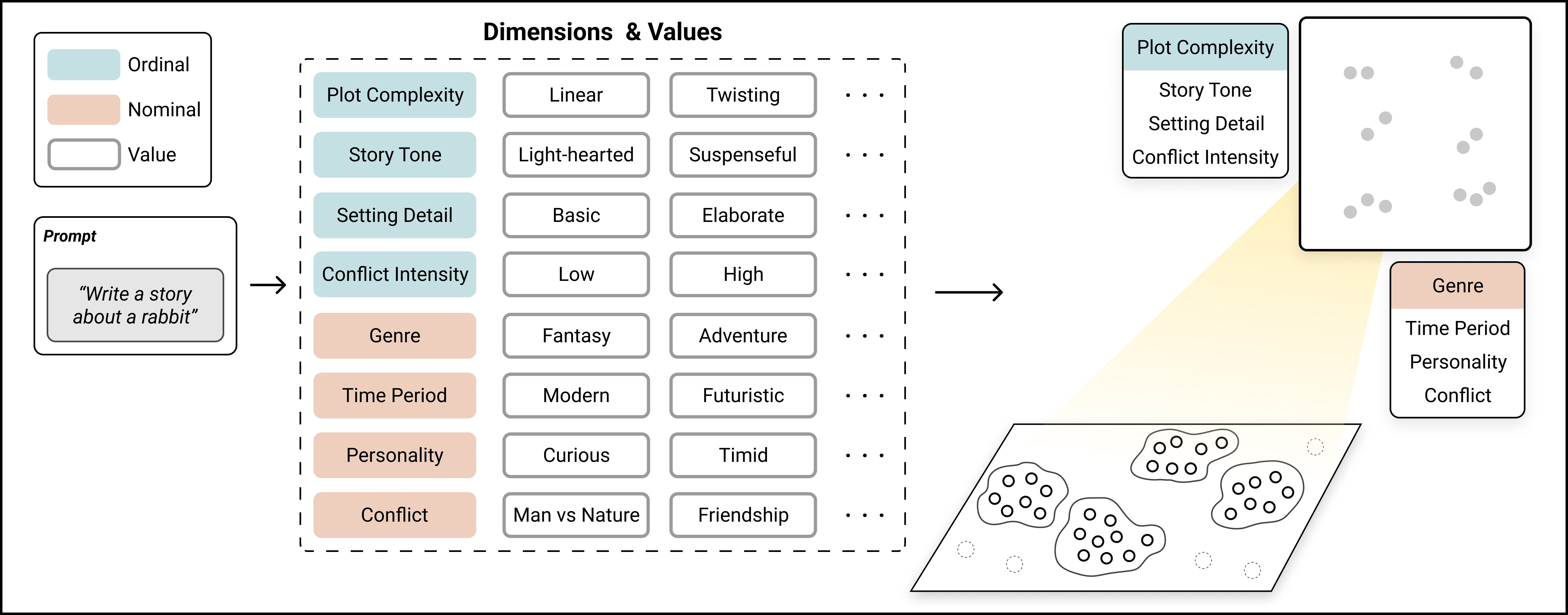}
    \caption{An example showing the benefits of exposing design space to users. By generating relevant dimensions for a task or topic, we can inform users about many \textsf{dimensions} and [\code{values}] they can consider --- e.g., \textsf{plot complexity}: [\code{linear}, \code{twisting}, ...], \textsf{genre}: [\code{fantasy}, \code{adventure}, ...]. If all the possible dimensions and values are used to generate responses, the responses will cover many sub-spaces within the overall design space and reveal to users a space of possible responses they can generate with LLMs.}
    \label{fig:framework-example}
\end{figure*}

\section{\textsf{Prompting for Design Space}: \\A Framework for Design Space Thinking in Human-AI Co-Creation}
\label{sec:framework}

Motivated by the benefits of design space thinking in creative processes --- (1) avoiding fixation~\cite{janis1982groupthink, jansson1991design} and (2) enabling systematic generation and evaluation of ideas~\cite{dove2016argument, macneil2017dimensional}, we argue that generative AI models should assist in the generation of the design space rather than individual artifacts at the early stage of the creative process to empower users and harness the creative potential of AI.

\subsection{Current Interaction Paradigms (Figs.~\ref{fig:teaser}A \& B)}
The way we interact with LLMs is designed to help users converge rather than diverge. Fig.~\ref{fig:teaser}A and Fig.~\ref{fig:teaser}B --- two prevalent interaction paradigms --- illustrate this. In Fig.~\ref{fig:teaser}A (unstructured single-output), users iteratively engineer prompts to generate a new output or refine the previous output. For instance, this would be equivalent to people asking ChatGPT to write a story about a rabbit and it returns one particular story; if they are not satisfied and want to refine or tweak the story, they will change (or refine) the prompt (Fig.~\ref{fig:teaser}a1) to find another version of the story. On the other hand, Fig.~\ref{fig:teaser}B (unstructured multi-output) represents an interaction where people ask the system to generate multiple responses (e.g., \raisebox{-2pt}{\includegraphics[scale=0.125]{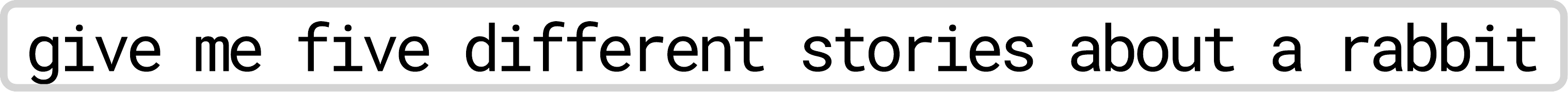}}, \raisebox{-2pt}{\includegraphics[scale=0.125]{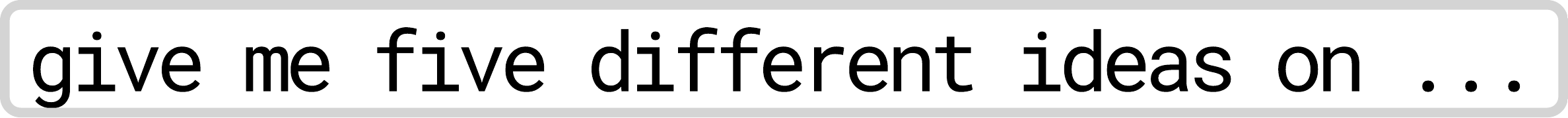}}) or the system generates multiple outputs for its users (e.g.,~\cite{brade2023promptify, feng2023promptmagician}).

In either case, if users' prompt is concise, missing enumeration of key details like \textsf{setting}, \textsf{plot}, \textsf{character personality}, \textsf{tone}, and so on, LLMs will fill in those missing details in their output. For example:

\begin{displayquote}
\textit{Once upon a time, in a peaceful \textbf{meadow} [\ul{setting}] nestled between rolling hills, there lived a small and \textbf{curious} [\ul{personality}] rabbit named Flora. Flora's fluffy, white fur and bright, inquisitive eyes made her stand out among her fellow rabbits. She was known for her \textbf{adventurous spirit} [\ul{personality}] ... Flora shared her stories of the underground world and the Heartbloom. From that day on, the \textbf{meadow} [\ul{setting}] became a place of wonder and adventure [\ul{tone}] for all the rabbits ...}
\end{displayquote}

Upon reading, users may \textit{realize} they do not wish for the character to possess curious or adventurous traits, nor desire the story to center around an underground adventure. They recognize the need for further specification of essential story elements, such as the \textsf{plot} and \textsf{character personality}. They refine their prompt, providing additional details: `Write a story about a rabbit named Jim, renowned in his village for his reserved personality. The plot of the story should revolve around...' As they iterate, they gradually uncover various aspects crucial to storytelling. However, as these aspects naturally emerge, they may not think to explore alternative ideas along those aspects and engage in divergent thinking. This constrained exploration can result in fixation on a single narrative direction, such as narratives that vary along certain dimensions (e.g, \textsf{setting}) but focus solely on a shy character. Consequently, this constrains the scope of creative exploration and leaves them stagnant in a small region of the design space. This highlights two issues: (1) \textbf{unstructured generation}: the outputs are generated in a haphazardous manner; (2) \textbf{unstructured exploration}: users' exploration of the design space is analogous to rummaging through the dark without a clear sense of direction and comprehensive view of the design landscape they are treading.

\subsection{Structured Multi-Output Approach (Fig.~\ref{fig:teaser}C)}
\label{sec:our-approach}

\subsubsection{Motivation}
\label{sec:framework_motivation}

This framework is motivated and grounded in research that uncovered (1) the challenges associated with dimensional reasoning despite its advantages and (2) the benefits of steering users toward parallel exploration of ideas in creative processes.

First, people normally do not reason with dimensions, mainly due to the challenges associated with identifying (and remembering) the list of relevant dimensions. MacNeil et al.~\cite{macneil2017dimensional} found that even experts struggle when asked to compile a list of relevant dimensions within their domains of expertise.  However, if available, they can serve as valuable tools for systematically comparing ideas and even catalyzing the generation of new ones. After all, dimensions serve as pathways for contemplating and exploring the design landscape. Since LLMs have shown an impressive capability, there are reasons to believe that they are also capable of recognizing and generating relevant dimensions for constructing appropriate design spaces.

Second, the literature on creativity consistently underscores our tendency to prematurely converge~\cite{janis1982groupthink, jansson1991design, cross2004expertise, dow2010parallel}, unless interventions are employed,  such as explicitly requiring the generation of multiple ideas at the start of the creative process~\cite{osborn1953applied, dow2010parallel}. For instance, prior work compared the effect of mandating the generation of multiple ideas before providing feedback with providing instant feedback on a single idea. They found that divergent ideation promoted comparison and exploration, and led to better results and self-efficacy~\cite{dow2010parallel}. This comparison between divergent and serial methods, drawn from prior work, bears a resemblance to the contrast between current interaction paradigms and our framework, as depicted in Fig.~\ref{fig:teaser}. The current interaction paradigms (Fig.~\ref{fig:teaser}A \& B) resemble the serial method, where users are encouraged to promptly evaluate the response. In contrast, our framework (Fig.~\ref{fig:teaser}C) initiates by generating dimensions and responses, thereby engaging users in reflective exploration of the design space before evaluation.

\subsubsection{Design Goals} 

Motivated by the challenges hindering dimensional thinking and creative processes, we propose the following design goals (\textbf{DG}s):

\begin{itemize}
    \item \textbf{DG1. Reduce the difficulty of generating dimensions by harnessing LLMs.} 
    \label{dg1}
    Although challenging even for experts, generating relevant dimensions for a specific topic, task, and domain can be effortlessly accomplished by LLMs, thanks to their extensive training on terabytes of data. Thus LLMs should be leveraged to ease the process of generating dimensions. 
    \item \textbf{DG2. Leverage dimensions to support structured generation of responses.} 
    \label{dg2}
    The goal of our framework is to not only help generate responses that reside in many sub-spaces within the design space but also do so in a structured, less haphazardous manner. By using dimensions, users can gain a better understanding of why and how certain responses were generated and what additional responses can be generated. Users should be able to use dimensions and their corresponding values to generate responses with those specific values.
    \item \textbf{DG3. Support flexible transition between divergent and convergent thinking.} 
    \label{dg4}
    While our framework places significant emphasis on the divergent phase of the creative process, it was developed with the understanding that it should seamlessly integrate into the entire creative process. Therefore, its implementation should enable a flexible and iterative interplay between divergent and convergent thinking. Users should be able to quickly transition from exploring ideas to evaluating specific ideas and vice versa, as needed. 
    \item \textbf{DG4. Enable users to engage in parallel exploration of ideas at every opportunity.}
    \label{dg3}
    As previously mentioned, parallel exploration of ideas can reduce fixation, lead to better and more diverse outcomes, and enhance self-efficacy~\cite{dow2010parallel}. Even when user is in the convergent phase, there should be no friction for users to switch to divergent phase and initiate parallel exploration of ideas. In fact, the system should encourage users to engage in creative exploration by initiating parallel idea exploration whenever users prompt it.
    \item \textbf{DG5. Support efficient and structured navigation of the design space.} 
    \label{dg5}
    Nonlinear canvases are more challenging to navigate than linear conversational interfaces~\cite{suh2023sensecape}, as users have a greater degree of freedom in their movement --- i.e., they can move not only vertically but also horizontally. However, given that the design space is represented as 2-dimensional, nonlinear space with ideas positioned within it, we should make it easy for users to locate, navigate to, and examine responses of interest.
\end{itemize}

\begin{figure*}[htp!]
    \centering
    \includegraphics[width=\textwidth]{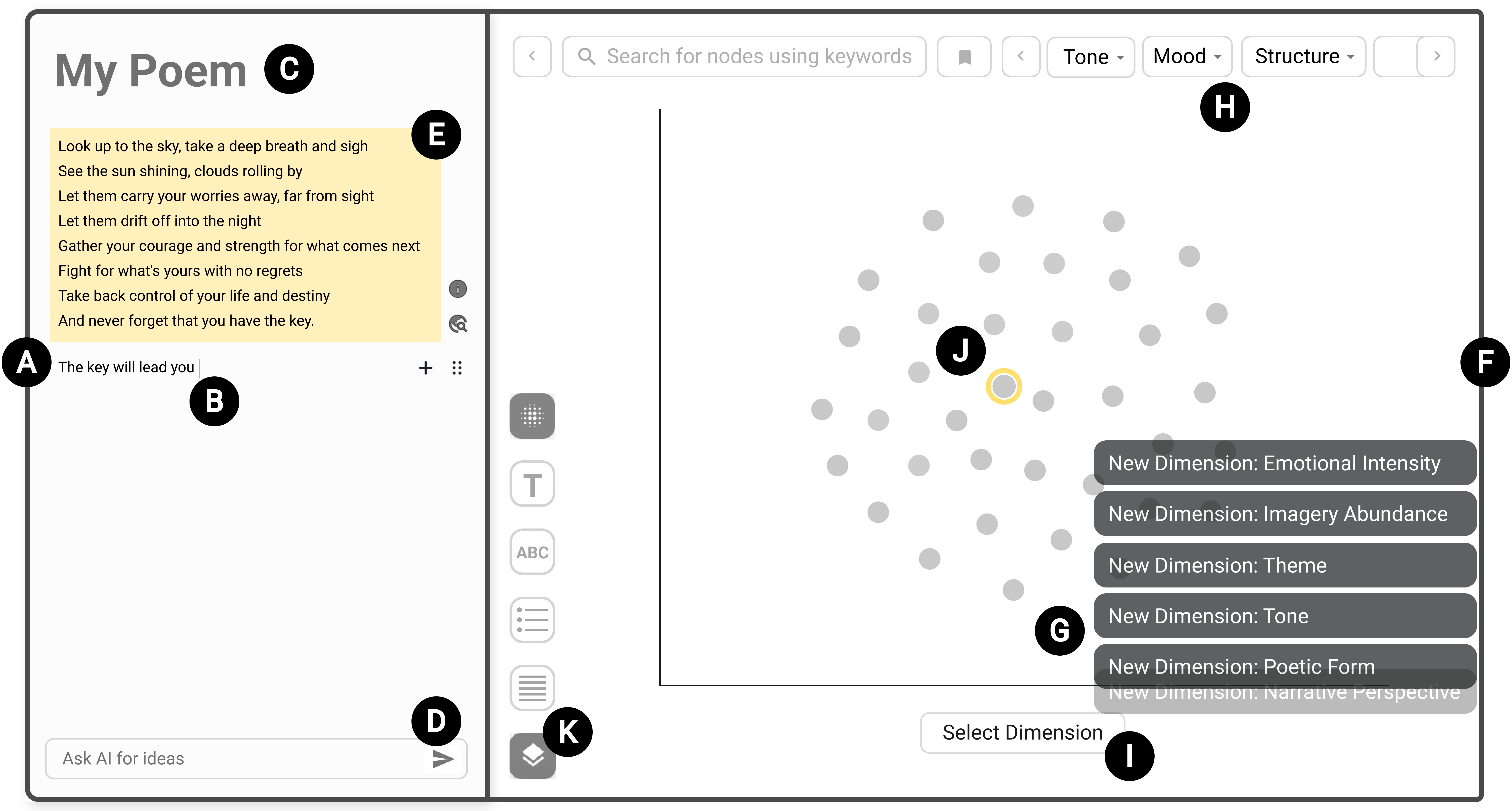}
    \caption{\sys{} interface consists of two main sections: (A) text editor and (F) exploration view. Users can (B) type and use various (C) text styles (e.g., \textsc{Title}, normal text). They have the option to (D) input a prompt (e.g., \raisebox{-2pt}{\includegraphics[scale=0.125]{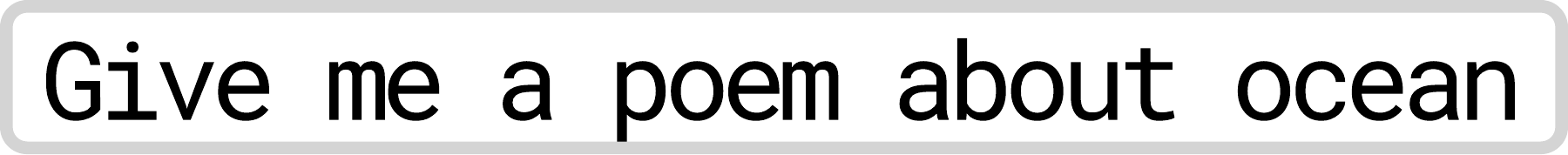}}) to LLM and view (E) one of the generated responses in the text editor. (F) In the exploration view, users can observe (J) a cluster of responses and (H) dimensions. At first, no dimension is selected, to invite users to explore the design space. Users can add dimensions to the axes (I) and arrange responses by their dimension values (see Fig.~\ref{fig:dimension-select}). For optimal experience, we promptly display (G) dimensions as soon as they are generated, allowing users to examine them while responses are being generated. Users can click (K) semantic level icons to adjust the zoom scale and view responses at different levels of detail (see Fig.~\ref{fig:semantic-zoom}).}
    \label{fig:interface}
\end{figure*}

\section{\sys{}}
\label{sec:system}

In this section, we first present an overview of the \sys{} interface, detailing its individual features, and demonstrate its workflow with a creative story writing task. 
Finally, we provide a concise overview of its implementation.

\subsection{Interface \& Features}
The \sys{} interface has two main sections: text editor (Fig.~\ref{fig:interface}A) and navigation canvas (Fig.~\ref{fig:interface}B). In the text editor, users can type and format text. They can also use the chat input box (Fig.~\ref{fig:interface}D) at the bottom to ask LLM for ideas. On the right side of the interface (Fig.~\ref{fig:interface}B) is the exploration view that displays generated responses. As shown, the responses are initially presented as a single cluster (Fig.~\ref{fig:interface}G).
When users select a node (\raisebox{-2pt}{\includegraphics[scale=0.1]{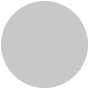}}), this adds the full text to the text editor and highlights it in yellow, as shown (Fig.~\ref{fig:interface}E). The selected node is also highlighted in yellow to indicate the mapping with text (Fig.~\ref{fig:interface}J) and show which node is currently selected.
To see other responses, users can click other nodes, replacing the text highlighted in yellow with the response the selected node matches. Once users edit this text, the yellow highlight disappears to indicate that it is now a part of the users' writing. We explain each feature and corresponding interactions next.

\begin{figure*}[htb!]
    \centering
    \includegraphics[trim=0cm 0cm 0cm 0cm, clip=true, width=\textwidth]{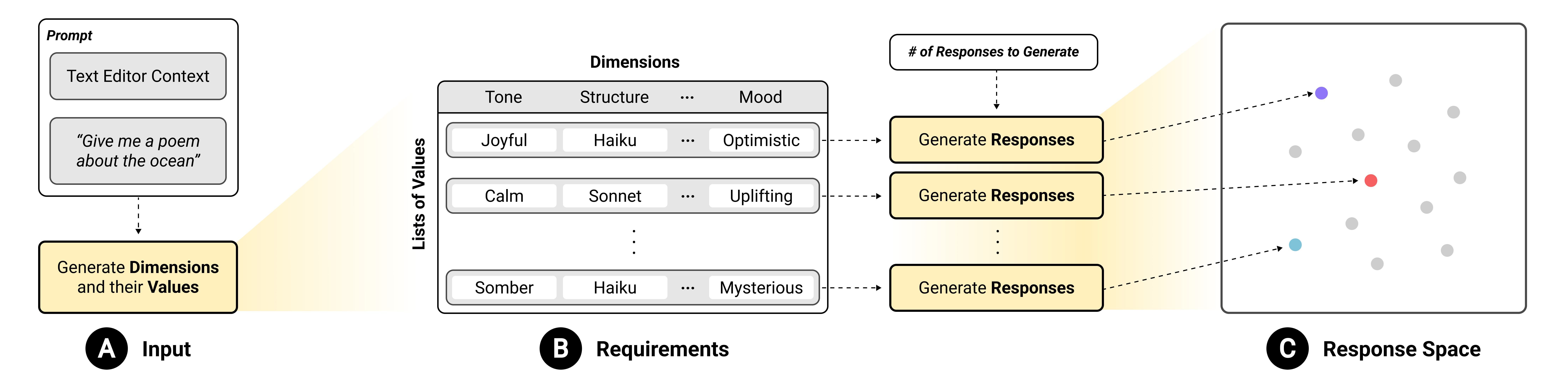}
    \caption{Technical pipeline of \textsf{Prompting for Design Space}: The pipeline consists of two LLM prompting steps. (A) The first LLM step feeds the inputs  from the text editor into the LLM API call to generate dimensions and their values as a JSON object. (B) \sys{} transforms the generated object to form a list of requirements per response by randomly selecting a value from each dimension. The second LLM step feeds each list into separate LLM API calls and generates all responses in parallel. (C) Generated responses are then visualized in \sys{}'s response space.}
    \label{fig:pipeline}
\end{figure*}

\subsubsection{Dimension Generation (Fig.~\ref{fig:pipeline}, \textbf{DG1})}

Users can initiate structured exploration by leveraging dimension generation in \sys{}. After users submit prompt (e.g., \raisebox{-2pt}{\includegraphics[scale=0.125]{figures/prompt/poem-idea.pdf}}) in the chat input box (Fig.~\ref{fig:interface}D), \sys{} takes that prompt and instructs an LLM to generate a pre-defined number (e.g., 5) of categorical dimensions and corresponding values (e.g., \textsf{Wave Intensity}: [\code{rolling waves}, \code{gentle ripples}]). To extract all possible dimensions, we instruct LLMs to generate ordinal and nominal dimensions for categorical dimensions. We exclude numerical dimensions such as \textsf{word length}, as our testing and pilot studies showed they do not deliver meaningful interactions and benefits.

\begin{figure*}[hbt!]
    \centering
    \includegraphics[width=\textwidth]{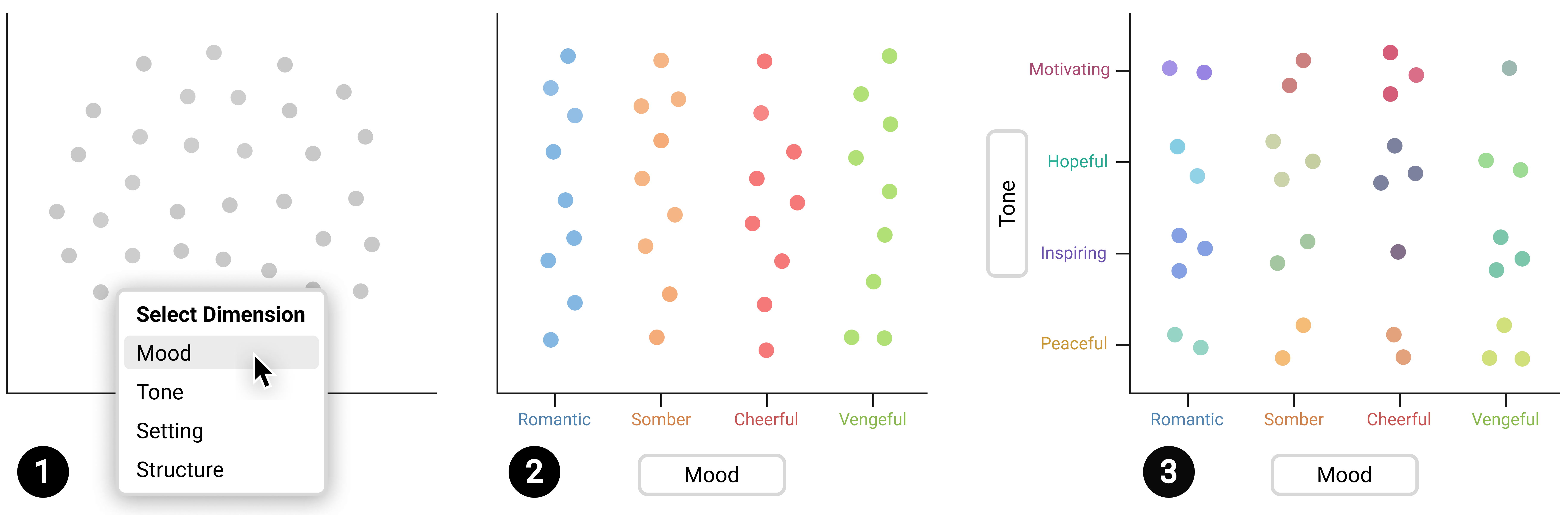}
    \caption{Dimension selection: upon (1) selecting a dimension (\textsf{Mood}) for the x-axis, responses (2) reposition to vertically align to their dimension values (\textcolor{ACMDarkBlue}{\code{Romantic}}, \textcolor{ACMOrange}{\code{Somber}}, \textcolor{ACMRed}{\code{Cheerful}}, \textcolor{ACMGreen}{\code{Vengeful}}). (3) Selecting a dimension for the y-axis (\textsf{Tone}) additionally repositions the responses to align horizontally to their dimension values (\textcolor{ACMPink}{\code{Motivating}}, \textcolor{ACMLightBlue}{\code{Hopeful}}, \textcolor{ACMPurple}{\code{Inspiring}}, \textcolor{ACMYellow}{\code{Peaceful}}).}
    \label{fig:dimension-select}
\end{figure*}

\subsubsection{Dimension-Guided Response Generation (Fig.~\ref{fig:pipeline}, \textbf{DG2})}
After dimensions are generated, \sys{} uses those dimensions and their values to guide the response generation. For example, suppose two of the generated \textsf{dimensions} and their [\code{values}] are: 
\textsf{Narrative Perspective}: [\code{first-person}, \code{third-person}] and \textsf{Poetic Style}: [\code{haiku}, \code{sonnet}, \code{free verse}]. 
\sys{} randomly selects a value for each dimension to create a \textit{requirement} --- a list of dimensions and their corresponding values (e.g., \textsf{Narrative Perspective}: [\code{first-person}], \textsf{Poetic Style}: [\code{haiku}]) --- and instructs the LLM to generate responses using these requirements, as illustrated in Fig.~\ref{fig:pipeline}. 
Fig.~\ref{fig:framework-example} and Table~\ref{table:dimension-prompts} in Appendix~\ref{sec:appendix} show few examples of the dimensions and responses generated for different creative writing tasks.

\subsubsection{Dimension Selection (Fig.~\ref{fig:dimension-select}, \textbf{DG5})}

Users can cluster and arrange generated responses by selecting dimensions. This allows them to quickly locate the right responses but also explore the design space in a structured manner. For example, as shown in Fig.~\ref{fig:dimension-select}, once users select a dimension (e.g., \textsf{Mood}) or two, the layout of the nodes changes, from (1) a cluster to (2) a one-dimensional or two-dimensional scatter plot.

\begin{figure*}[htb!]
    \centering
    \includegraphics[trim=0cm 0cm 0cm 0cm, clip=true, width=\textwidth]{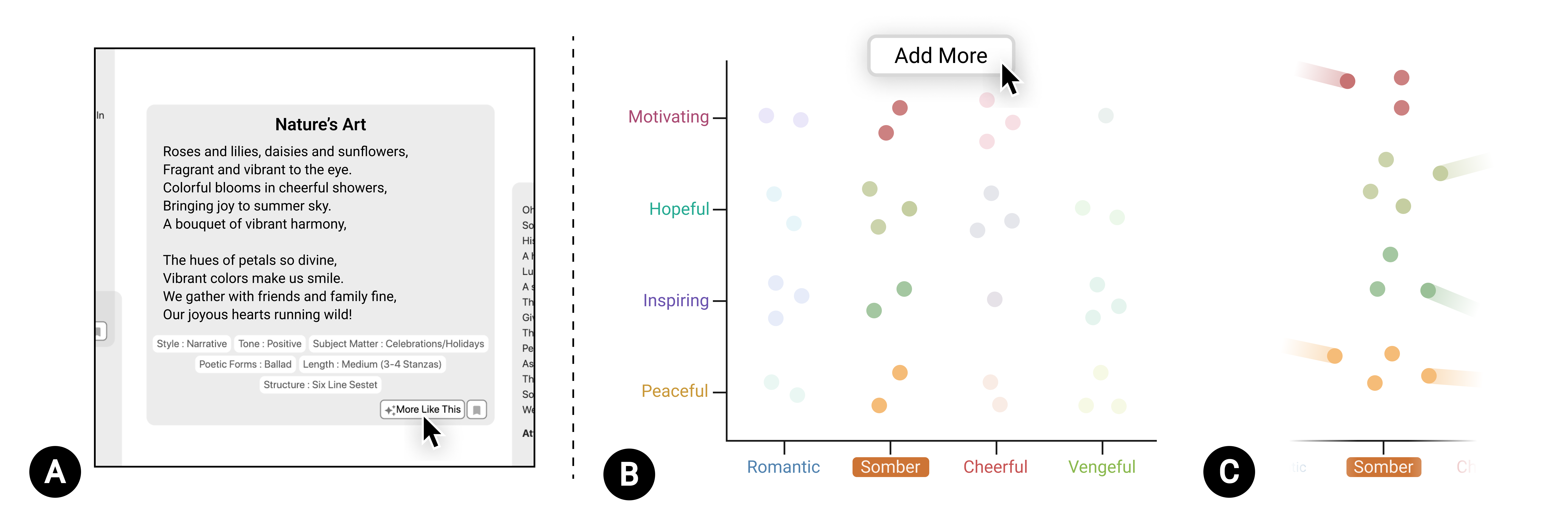}
    \caption{Two methods for generating similar responses: Method\#1: (A) Users can click the \raisebox{-2pt}{\includegraphics[scale=0.19]{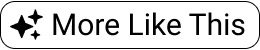}} button on the node to generate similar responses. Method\#2: (B) When nodes are filtered on the canvas, the user can click the \raisebox{-2pt}{\includegraphics[scale=0.19]{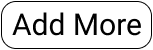}} button to generate more nodes to be added to the filtered subspace; (C) newly generated nodes join the canvas and align themselves to their dimension values.}
    \label{fig:space-regeneration}
\end{figure*}

\begin{figure}[htb!]
    \centering
    \includegraphics[trim=0cm 0cm 0cm 0cm, clip=true, width=0.85\columnwidth]{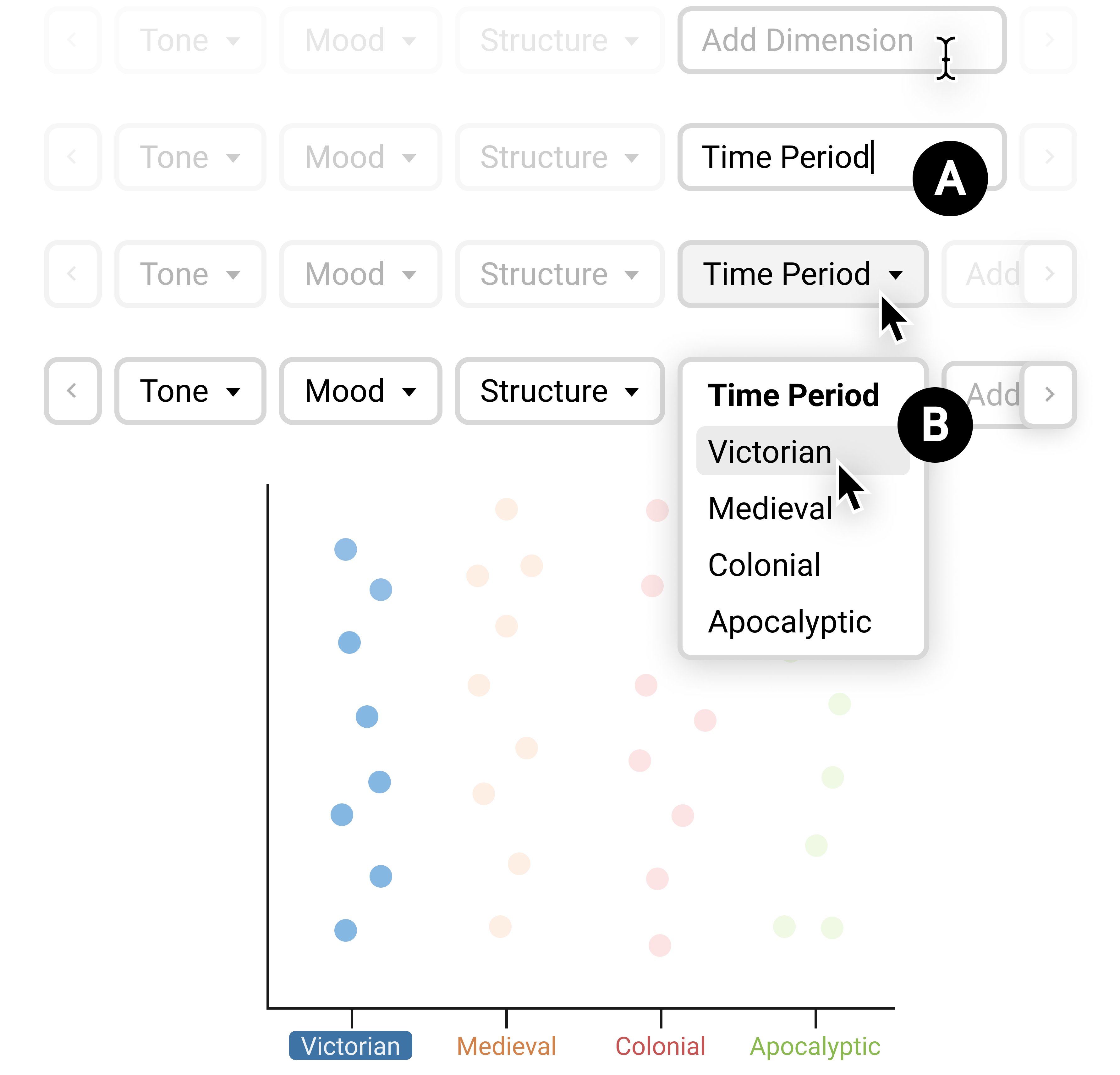}
    \caption{Generating new responses via user-defined dimension: Users can generate new responses by (A) adding new dimensions to their exploration space. Upon entering their desired dimension in the filter bar, \sys{} generates (B) values (e.g., \textcolor{ACMDarkBlue}{\code{Victorian}}, \textcolor{ACMOrange}{\code{Medieval}}, \textcolor{ACMRed}{\code{Colonial}}, \textcolor{ACMGreen}{\code{Apocalyptic}}) for the desired dimension ( \textsf{Time Period} ) and revises all existing responses to include one of the value for that response. Once complete, users can use the new dimension to search, filter, and cluster the revised responses.}
    \label{fig:add-dimension}
\end{figure}

\subsubsection{Generation of New Responses (Fig.~\ref{fig:space-regeneration} \& \ref{fig:add-dimension}, \textbf{DG4})}
\label{sec:generate_new_results}

\sys{} offers three novel techniques for steering the generation of new responses. Specifically, users can add additional responses or add additional attribute(s) to existing responses.
The first two support the generation of responses with specific attributes within existing dimension(s). Users can select the \raisebox{-2pt}{\includegraphics[scale=0.19]{figures/button/more-like-this.png}} button on the node to generate responses containing the same attributes (Fig.~\ref{fig:space-regeneration}A). Or they can first select label(s) to specify a subspace in the design space and then the \raisebox{-2pt}{\includegraphics[scale=0.19]{figures/button/add-more.png}} button to generate new responses in the subspace, as shown in Fig.~\ref{fig:space-regeneration}B-C. The last technique is adding a user-defined dimension. As shown in Fig.~\ref{fig:add-dimension}, after users add a new dimension, \sys{} generates its values and then updates existing responses with new requirements (Fig.~\ref{fig:pipeline}B) containing these values.

\begin{figure*}[htb!]
    \centering
    \includegraphics[trim=0cm 0cm 0cm 0cm, clip=true, width=\textwidth]{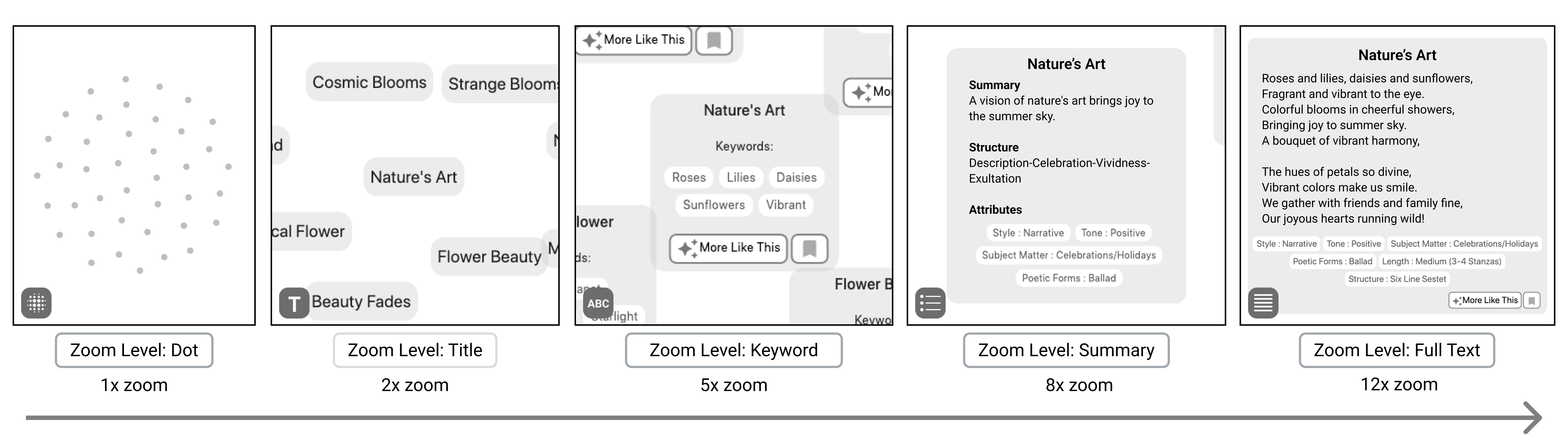}
    \caption{Semantic zoom: The amount of information users see changes with zoom scale. As shown, as users zoom in, the displayed information transitions from \raisebox{-1.6pt}{\includegraphics[scale=0.15]{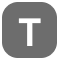}} \textsc{title} to \raisebox{-1.6pt}{\includegraphics[scale=0.15]{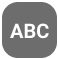}} \textsc{keyword}, then to \raisebox{-1.6pt}{\includegraphics[scale=0.15]{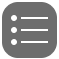}} \textsc{summary} and to \raisebox{-1.6pt}{\includegraphics[scale=0.15]{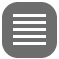}} \textsc{full text}. Users can double click anywhere on canvas to zoom directly into the \raisebox{-1.6pt}{\includegraphics[scale=0.15]{figures/icon/full-text.png}} \textsc{full text} level or zoom out to the \raisebox{-1.6pt}{\includegraphics[scale=0.15]{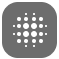}} \textsc{dot} level from the \raisebox{-1.6pt}{\includegraphics[scale=0.15]{figures/icon/full-text.png}} \textsc{full text} level, for efficient navigation of the space.}
    \label{fig:semantic-zoom}
\end{figure*}

\subsubsection{Semantic Zoom (Fig.~\ref{fig:semantic-zoom}, \textbf{DG5})}

\begin{figure*}[htb!]
    \centering

    \begin{subfigure}[t]{0.45\textwidth}
    \includegraphics[trim=0cm 0cm 0cm 0cm, clip=true, width=0.8\columnwidth]{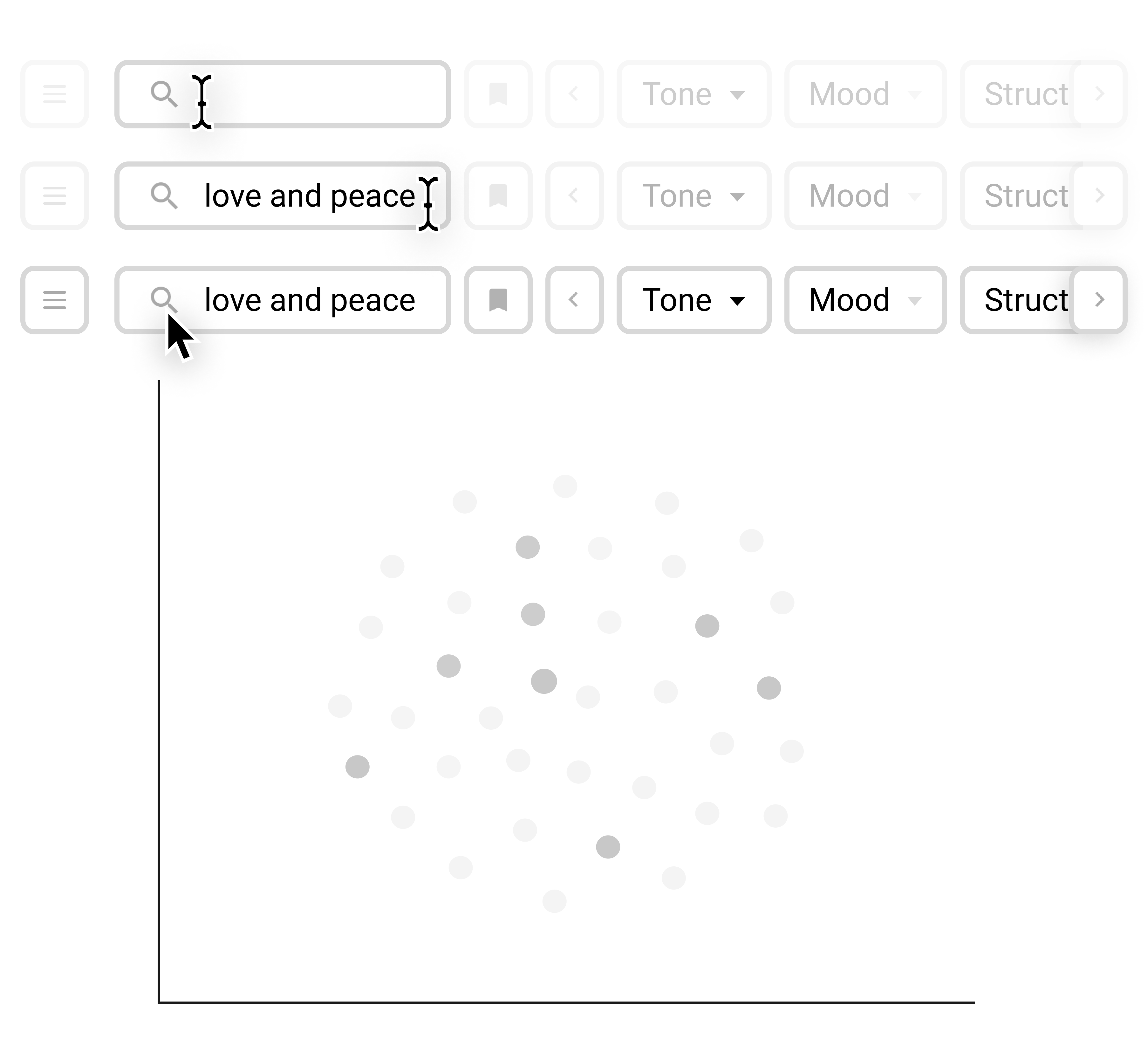}
    \caption{Search by Keyword}
    \label{fig:search-keywords}
    \end{subfigure}
    \begin{subfigure}[t]{0.45\textwidth}
        \includegraphics[trim=0cm 0cm 0cm 0cm, clip=true, width=\textwidth]{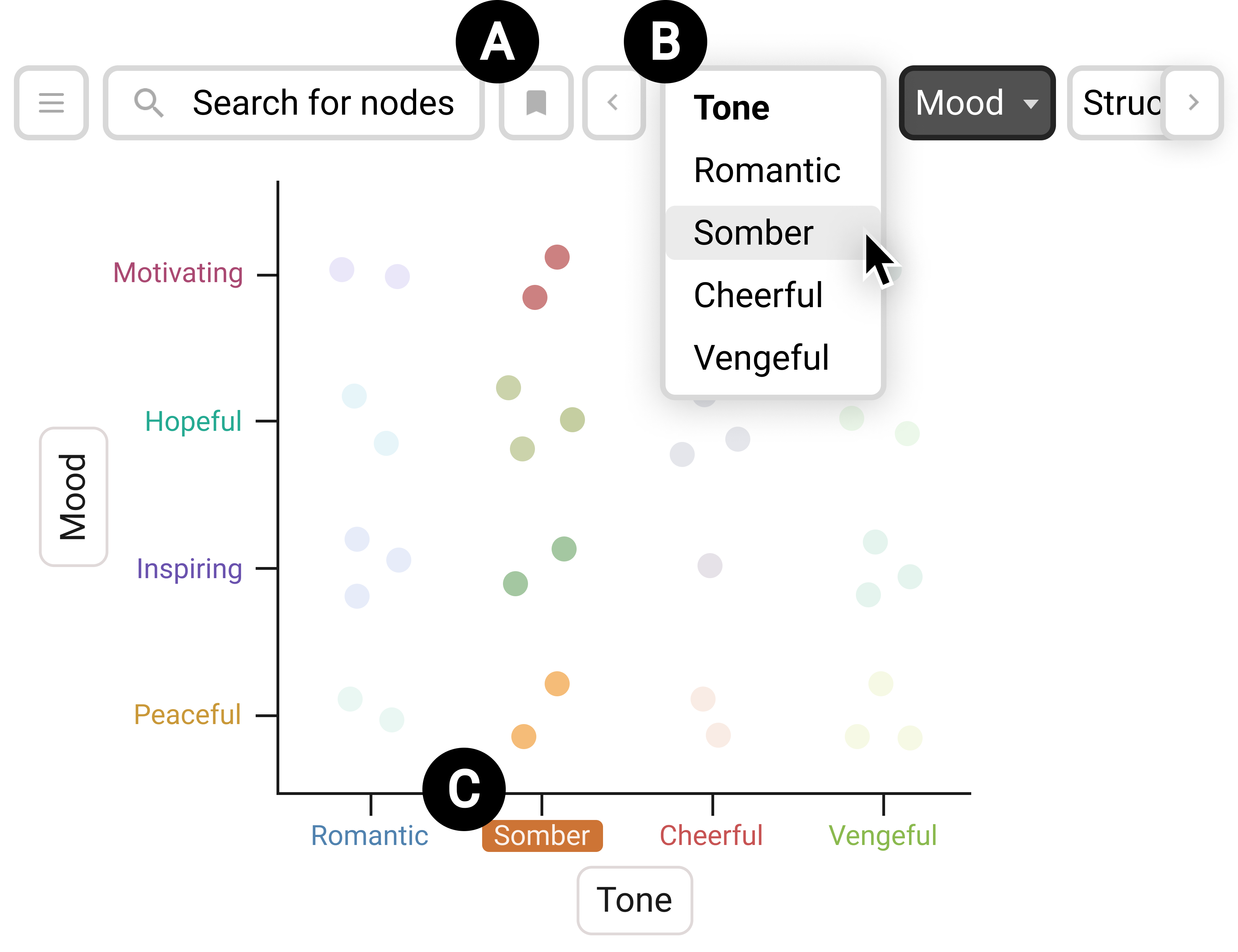}
        \caption{Search by Filter}
        \label{fig:search-filter}
    \end{subfigure}
    \caption{Search features: To address the challenge of \textit{locating} the desired response with specific properties, \sys{} offers two search methods. With (a), users can search for responses containing specific words such as `\textsf{love and peace}.' As shown, instead of completely removing the non-corresponding responses, we reduce their opacity, to preserve their spatial information while making the corresponding responses stand out. With (b), users can filter for (A) bookmarked nodes or by (B) dimension values using the dropdown. Alternatively, as shown, users can also (C) click on the dimension value ( \raisebox{-0.8pt}{\includegraphics[scale=0.19]{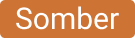}} ) to apply the filter.}
    \label{fig:search-features}
\end{figure*}

Motivated by recent work that demonstrated the benefits of offering users the ability to control the level of detail for LLM-generated text~\cite{dang2022beyond, suh2023sensecape, bederson1994pad++}, \sys{} uses \textit{semantic zoom} to give users control over the granularity of detail, facilitating their navigation within the subspace. As shown in Fig.~\ref{fig:semantic-zoom}, there are five levels of detail: \raisebox{-1.6pt}{\includegraphics[scale=0.15]{figures/icon/dots.png}} \textsc{dot}, \raisebox{-1.6pt}{\includegraphics[scale=0.15]{figures/icon/title.png}} \textsc{title}, \raisebox{-1.6pt}{\includegraphics[scale=0.15]{figures/icon/keywords.png}} \textsc{keyword}, \raisebox{-1.6pt}{\includegraphics[scale=0.15]{figures/icon/summaries.png}} \textsc{summary}, and \raisebox{-1.6pt}{\includegraphics[scale=0.15]{figures/icon/full-text.png}} \textsc{full text}. Users can manually zoom in and out (by scrolling on the mouse or pinch-to-zoom) or click the semantic level icons (Fig.~\ref{fig:interface}K) to adjust the view to the right level of detail. Users can also double-click on an empty space in the canvas to fully zoom in to view the full response. Double-clicking from the \raisebox{-1.6pt}{\includegraphics[scale=0.15]{figures/icon/full-text.png}} \textsc{full text} level brings them back to the \raisebox{-1.6pt}{\includegraphics[scale=0.15]{figures/icon/dots.png}} \textsc{dot} level.

\subsubsection{Search \& Filter (Fig.~\ref{fig:search-features}, \textbf{DG5})}

Users can use the search box to search for responses containing certain keywords. Once users type a keyword into the search box, \sys{} \textit{hides} responses that do \textit{not} contain those keywords by lowering their opacity. Users can also use the filter bar to filter responses by dimension values. These search features help users quickly find responses of interest.

\begin{figure*}[htb!]
    \centering
    \includegraphics[trim=0cm 0cm 0cm 0cm, clip=true, width=\textwidth]{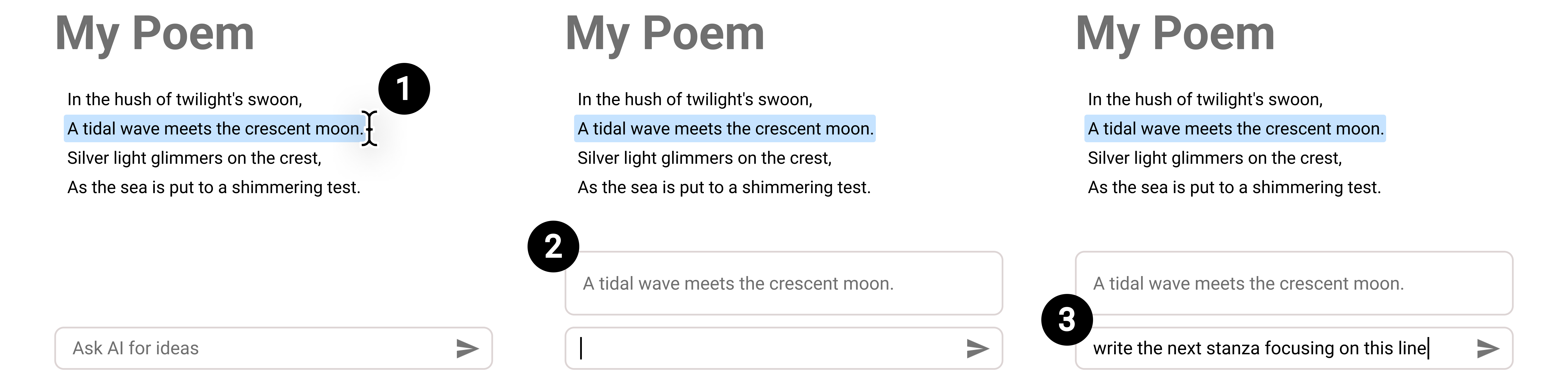}
    \caption{Contextual generation: Users can select text in the editor (1) to highlight a line or phrase (e.g., 
    \textsf{A tidal wave meets the crescent moon})
    to focus on as context. Users can then (2 \& 3) prompt \sys{} to generate responses given the focused context. This allows users to have a more fine-grained control over the generation of responses.}
    \label{fig:contextual-generation}
\end{figure*}

\subsubsection{Contextual Generation (Fig.~\ref{fig:contextual-generation}, \textbf{DG3-4})}
During our pilot studies, we found participants desired to generate responses that align with the writing they had added. To support this, \sys{} provides the LLM with the content present in the text editor as background context. However, there can be cases where users may want a more targeted generation --- e.g., expand on specific parts of the content as opposed to follow up on the entire content. Thus \sys{} allows users to highlight text in the editor when prompting for responses. We add the highlighted content into the prompt along with any other content present in the text editor (see Table ~\ref{table:prompt-list}). This provides users with more contextualized responses and a way to precisely steer the generation of responses.

\begin{figure*}[htb!]
    \centering
    \includegraphics[trim=0cm 0cm 0cm 0cm, clip=true, width=\textwidth]{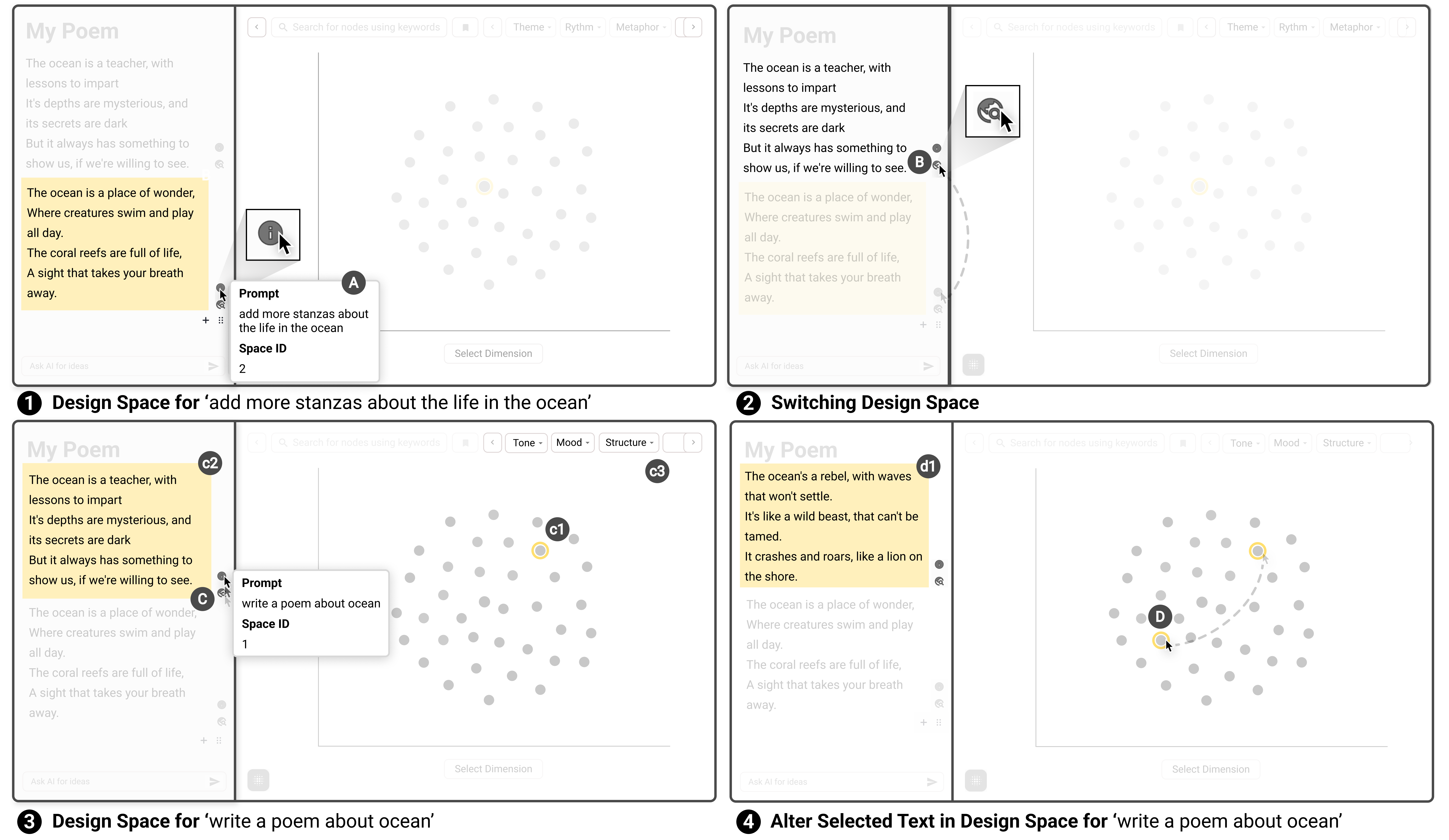}
    \caption{Switching to other design space: (A) Clicking the \textit{Show Information} (\raisebox{-1.6pt}{\includegraphics[scale=0.12]{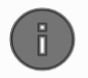}}) button displays the current design space prompt (`add one more stanza about life in the ocean') along with its unique space ID. (B) The \textit{Show Space} (\raisebox{-1.6pt}{\includegraphics[scale=0.12]{figures/icon/show-space.png}}) button allows users to switch back to the design space for prompt `write a poem about ocean'. (C) When the design space switches, the exploration view is updated, to show a cluster of responses, along with (c1) a node that corresponds to the (c2) text block in the editor highlighted, and (c3) dimensions in this design space (Space ID = 2). (D) Clicking on another node updates the content in the (d1) block to show its full text.}
    \label{fig:switching-space}
\end{figure*}

\subsubsection{Design Space Exploration (Fig.~\ref{fig:switching-space}, \textbf{DG4})}
When users engage in creative tasks, it is natural for them to go back and forth between a set of ideas. Thus, \sys{} supports a space-switching feature that enables users to change their selected text and explore different design spaces, which are sets of responses from different prompts, at anytime they want. As shown in Fig. \ref{fig:switching-space}, for example, a user has recently added a new stanza to the ocean-themed poem. In order to enhance the alignment of length and lexical level between the previously generated stanza and the newly added one, the user clicks the \textit{show space} (\raisebox{-1.6pt}{\includegraphics[scale=0.12]{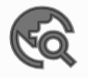}}) button and searches for an alternative stanza.

\subsection{Example Workflow: Writing Short Story}
Below we present an example workflow to demonstrate some of the features described above. Chris is a professional writer who writes children's books. He is attending a workshop for writers interested in learning about ways AI tools can help them in their creative writing process. The workshop provides Chris access to \sys{} and suggests he write a short story over the weekend and share it with other writers on Monday.

\subsubsection{Divergent Thinking Phase (Exploration)}
Chris opens \sys{} on the web browser. He is greeted by the \sys{} interface (Fig.~\ref{fig:interface}), with text editor on one side and exploration view on the other side. The text editor and exploration view are both empty. He writes a header `My Story' at the top of the text editor. Without any specific idea on what he wants to write about, he first recalls recent events and encounters for inspiration. He remembers running into a rabbit outside his house in the morning. Unsure what interesting story he can write about, he decides to ask AI. He writes in the chat input box: \raisebox{-2pt}{\includegraphics[scale=0.125]{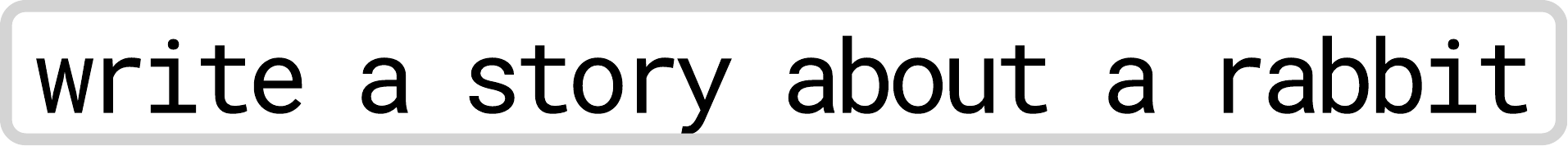}}. \sys{} starts to generate 30 stories. After about 5 seconds, \sys{} shows several toast messages (e.g., Fig.~\ref{fig:interface}G), with each specifying what dimensions LLMs found relevant: \textsf{plot}, \textsf{setting}, \textsf{genre}, \textsf{character type}, \textsf{tone}, and \textsf{originality}. Then after another 3 seconds, he sees a story highlighted in yellow in the text editor and a cluster of dots in the exploration view.

\subsubsection{Convergent Thinking Phase (Evaluation)}
Chris selects several dots randomly, which updates the text in the text editor, showing him the full text for each dot. After browsing a few and reading only the first one or two sentences, he zooms in further, which turns the dots into ellipses with titles of the story. They read: \raisebox{-2pt}{\includegraphics[scale=0.125]{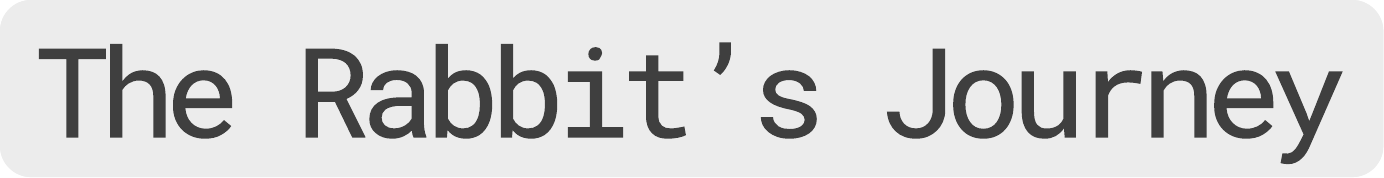}}\hspace{0.01in}, \hspace{0.01in}\raisebox{-2pt}{\includegraphics[scale=0.125]{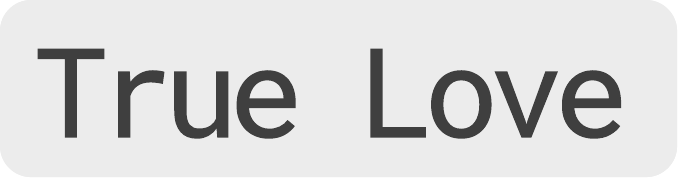}}\hspace{0.01in}, \raisebox{-2pt}{\includegraphics[scale=0.125]{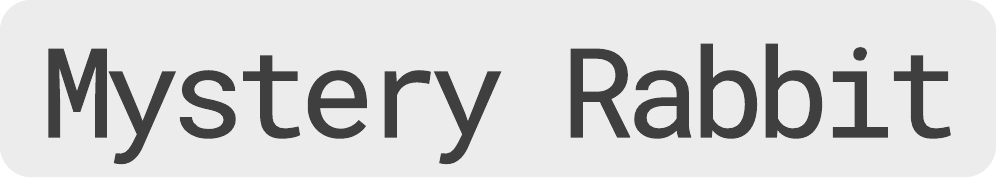}}\hspace{0.01in}, and so on. \raisebox{-2pt}{\includegraphics[scale=0.125]{figures/title/mystery-rabbit.pdf}} triggers his curiosity. He zooms in further to see keywords of the \raisebox{-2pt}{\includegraphics[scale=0.125]{figures/title/mystery-rabbit.pdf}} story\hspace{0.01in}, which reads: \hspace{0.02in}\raisebox{-2pt}{\includegraphics[scale=0.125]{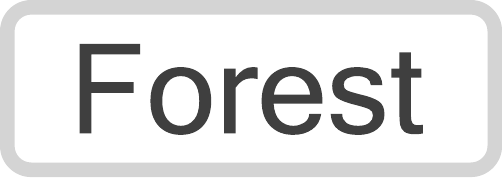}}\hspace{0.03in}, \hspace{0.05in}\raisebox{-2pt}{\includegraphics[scale=0.125]{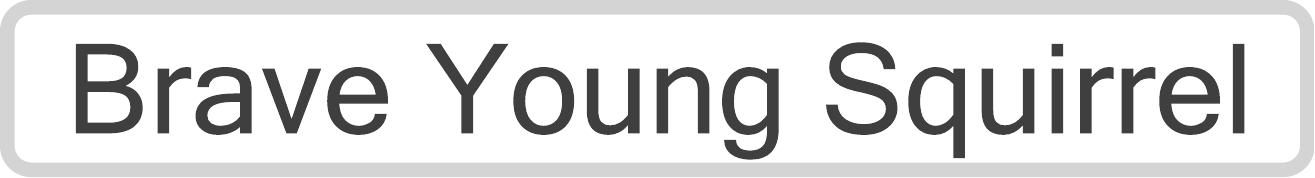}}, \hspace{0.02in}\raisebox{-2pt}{\includegraphics[scale=0.125]{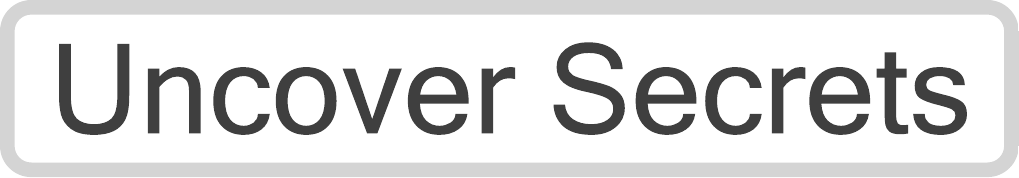}}.
Zooming in further reveals the summary of the story: `A Mysterious rabbit lurks in a forest, and a brave young squirrel sets off to uncover its secrets.' He realizes the protagonist of this story may be a squirrel, and the antagonist a rabbit. He had not considered this but realizes this might be an interesting direction he can take as well. The attributes listed below the summary confirm this, as it shows the \textsf{Character Type}: \code{Antagonist} tag. The other attributes --- \textsf{Plot}: \code{Mystery}, \textsf{Setting}: \code{Forest}, \textsf{Genre}: \code{Fantasy}, \textsf{Tone}: \code{Mysterious/Suspenseful} --- inform him what this story is about but also make him think about what other genres or settings he could explore. Although he can select the node and read the full text in the editor, he decides to zoom in to reveal the full text in the node. It reads:

\begin{displayquote}
\textit{Once upon a time, in a mystical forest, there lived a mysterious rabbit... The rabbit seemed to be up to no good, and his motives were unclear... One day, a brave young squirrel decided to confront him and ask what his mission was. Before she could get her answer, however, the rabbit disappeared into thin air! The squirrel decided to find out more about this mysterious creature and set off on an adventure to uncover his secrets...}
\end{displayquote}

Chris thinks this could be one starting point, but he is also curious if he can explore more stories like this one. He clicks a \raisebox{-2pt}{\includegraphics[scale=0.19]{figures/button/more-like-this.png}} button. \sys{} shows \raisebox{-2pt}{\includegraphics[scale=0.19]{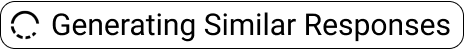}} and after 3 seconds, adds 5 more stories into the space.

\subsubsection{Divergent Thinking Phase (Exploration)} He zooms out to the \textsc{title} level to quickly grasp what the stories are about. He reads the titles of these new stories. The title for one of them reads, \raisebox{-2pt}{\includegraphics[scale=0.125]{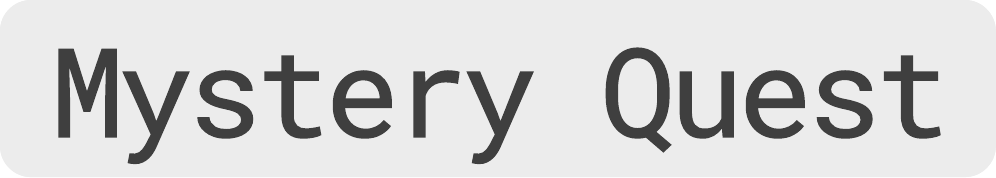}}, and the summary goes: `A white rabbit embarks on a journey to uncover the truth behind mysterious disappearances in the forest.' While interesting, he is not ready to commit to this direction, because he knows he has not yet explored all the dimensions. So he clicks the bookmark icon (\raisebox{-2pt}{\includegraphics[scale=0.12]{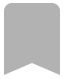}}) to save the story. Then he decides to explore more broadly by manipulating dimensions. He clicks the \raisebox{-2pt}{\includegraphics[scale=0.19]{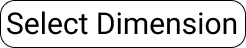}} button and selects \textsf{Genre}. The labels representing values along the \textsf{Genre} dimension --- \textcolor{ACMDarkBlue}{\code{Mystery}}, \textcolor{ACMOrange}{\code{Fairytale/Folklore}}, \textcolor{ACMBlue}{\code{Romance}}, \textcolor{ACMPurple}{\code{Sci-fi/fantasy}}, \textcolor{ACMRed}{\code{Horror}}, and \textcolor{ACMGreen}{\code{Adventure}} --- are added to the x-axis. The nodes that were in a single cluster moved to align with the corresponding labels. Intrigued by how he can think and explore in terms of dimensions, he selects another dimension, \textsf{Setting}, which replaces previous labels in the x-axis with labels such as \textcolor{ACMPurple}{\code{forest}}, \textcolor{ACMOrange}{\code{urban area}}, \textcolor{ACMRed}{\code{underground}}, and \textcolor{ACMDarkBlue}{\code{underwater
}}. He is surprised to see \textcolor{ACMDarkBlue}{\code{underwater}} and zooms into the \textsc{summary} level to see some interesting stories about a rabbit named Tony in an underwater kingdom. He bookmarks the story, zooms out, and explores a few other dimensions as well. 

\subsubsection{Convergent Thinking (Synthesis)} Eventually, he feels he has collected enough ideas to write a story. He clicks a bookmark icon (\raisebox{-2pt}{\includegraphics[scale=0.12]{figures/icon/bookmark.png}}) to find the stories he saved --- some for the phrases or sentences he particularly liked and others he wanted to use as potential ideas for aspects of his story such as setting, conflict, and tone. He clicks the nodes to display the text in the text editor and edits it, which embeds that text block into the text editor. When he selects another node, it appends the text under the previously embedded text. Chris goes on to edit the text to write a draft of his story.

The scenario above demonstrates a flexible and iterative interplay between divergent and convergent thinking. This workflow, however, is one of numerous potential workflows and does not encompass all the features within \sys{}.

\subsection{Implementation \& Prompt Engineering}
\sys{} was implemented using React, Editor.js, and d3.js. The text editor was implemented using Editor.js, and the exploration view and animation effects (clustering and positioning of points) with d3.js. Specifically, we used the force layout mechanism in d3.js for the clustering of points.

We used OpenAI's text-davinci-003 API to generate dimensions, dimension values, and responses.
Given the user's prompt and context (Fig.~\ref{fig:contextual-generation}), \sys{} first prompts the LLM to generate a set of categorical (ordinal and nominal) dimensions followed by their dimensional values (see Table~\ref{table:dimension-prompts}).
Once dimensions and their values are generated, \sys{} then forms requirement lists (e.g.,~\cite{kim2023metaphorian}) with one random value from each dimension, and for each requirement list, prompts the LLM to generate responses that fall under those requirements. Once each response is generated, \sys{} additionally prompts the LLM to generate abstractions of the response for each semantic level (Fig.~\ref{fig:semantic-zoom}).
All subsequent generations of new responses repeat the process of generating new responses but form requirement lists from the filtered dimension values instead. All prompts used in this pipeline are listed in Table \ref{table:prompt-constant} - \ref{table:dimension-prompts} in Appendix~\ref{sec:appendix}. 

\section{User Study}

To investigate the feasibility and usefulness of our framework, we conducted a user study with \sys{}. We recruited professional writers with experience using AI tools for their creative writing tasks in order to understand how \sys{} compares with existing AI creativity support tools.
The screen and audio were recorded for accurate transcription and analysis. For their participation in an 80-minute study, they received 40 USD. 

\subsection{Participants} 
Fourteen professional writers (9F, 5M; P1-P14; M$_{age}$= 30, SD = 7.7) were recruited via Upwork~\cite{upwork}. 
All of them had several years of experience (M = 7.3; SD = 8.4, range = [2, 35]) in creative writing, had experience incorporating AI-generated writing into their writing, and had exposure to various AI tools (14 ChatGPT, 4 JasperAI, 4 DALL-E, 1 Stable Diffusion, 1 frase.io, 1 QUILLBOT, 1 Bard, 1 Sage Poe, 1 Claude). They shared that approximately 26\% of their final writing was AI-generated (SD = 15.4\%). Most of them frequently used ChatGPT (6 on a daily basis; 6 on a weekly basis; 2 on a monthly basis). They varied in terms of how they used AI tools. Almost all of them (13/14) had used them for idea generation (e.g., generate creative ideas, story concepts), 11 for editing and proofreading (e.g., grammar and style checks), 7 for assistance in descriptive writing (e.g., add descriptions of settings, scenes, or characters), and 5 for content expansion (e.g., expand on existing content), 5 for world-building (e.g., create worlds with cultures and settings, for use in novels, games, or other creative projects), 5 for getting suggestions on character names and book and chapter titles, and 1 for creating different versions of the writing.

All participants had experience engaging in more than one creative writing task professionally. All 14 participants had experience in copywriting. Eleven had written a story and 10 an email (e.g., marketing email); 6 had experience in writing a letter (e.g., cover letter) and 4 a poem; there was one participant with experience in various tasks, including screenwriting, how-to guides, travel guides, product reviews, blog writing in their respective professional experience, composing song lyrics, writing story outlines for video games, comics, article, character development, and sci-fi stories. To hear them compare their experience using \sys{} with other AI tools they have previously used and investigate the scalability of this approach, we assigned creative writing tasks based on their prior experiences: 4 participants were asked to do copywriting, 4 crafting a short story, 2 writing an email, 2 writing a poem, 1 constructing video game scene, and 1 composing song lyrics.

\subsection{Procedure}
\label{sec:procedure}

\noindent\textit{Set-up and Pre-study Survey (10 min).} All 14 studies were hosted online on Zoom. After receiving participants' signatures on the consent forms and their approval of being recorded video and audio for later transcription and review, the study started off with a pre-study questionnaire to collect their demographic information (i.e., name, age, gender), experience in creative writing (i.e., years, types), and previous exposure and experience using AI tools in creative writing (i.e., frequency of usage, kinds of AI tools, type of creative writing tasks). 
\newline

\noindent\textit{Interface Tutorial (25 min).} 
After completing the pre-study survey and receiving the link to \sys{}, participants shared their screen and granted the researcher remote control, which enabled the latter to demonstrate how to use the system. In the tutorial session, researchers first introduced the interface to the user and then showed interactions via a practice task --- \raisebox{-2pt}{\includegraphics[scale=0.125]{figures/prompt/rabbit-idea.pdf}}. The researcher explained the purpose and functionality of each interaction and demonstrated a few actions at a time, followed by users performing the same actions as practice. This ensured participants had enough practice prior to the actual task and could ask the researcher if anything was unclear. The researcher followed the script containing detailed instructions, to ensure all participants thoroughly understood all the features in the system, and undergo the same tutorial.
\newline

\noindent\textit{Creative Writing Task with AI Support (25 min).} 
Once they were familiar with the system, participants had up to 25 minutes to use the system for a creative writing task. Participants performed think aloud throughout the whole section.
Participants could choose any topic they want in a category decided by the researcher based on their previous experience (see Table~\ref{table:user-study-topic} for the list of topics and writing tasks).
During the study, \sys{} was configured to generate up to 5 nominal dimensions with up to 8 different labels and up to 3 ordinal dimensions with 5 levels (\code{least}, \code{less}, \code{neutral}, \code{more}, \code{most}). 
Based on the feedback from our pilot studies, we limited the length of LLM-generated responses to 150 words and set the number of responses per prompt to 40, to avoid overwhelming participants but provide enough variations for creative exploration.
\newline

\noindent\textit{Post-study Survey and Exit Interview (20 min).} 
Following the completion of the task, a post-study survey was administered. It contained: the Creativity Support Index (CSI) questionnaire, questions related to the quality of AI-generated dimensions, and the evaluation of system's usability and usefulness of individual features. The study ended with a semi-structured interview in which users provided qualitative feedback on the system regarding the quality of the dimensions, compatibility with their workflow, usability, and effectiveness of features. For example, we asked them to compare \sys{} with other AI tools and how they might integrate \sys{} into their workflow.

\section{Results}

We present analysis of survey responses, participants' interactions with \sys{} (e.g., how many responses they explored), observation, and interview, describing participants' general assessment of \sys{}, their workflows, as well as how \sys{} helps steer users toward divergent thinking and enables an understanding of the design space.
\begin{table}[ht]
\centering
\caption{Creativity Support Index (CSI) Results (N=14). The highest value is in \textbf{bold}. The second highest in \underline{underline}. Since our study did not involve collaboration, we followed the practice from \cite{carroll2009creativity, suh2022codetoon}, omitting the \textit{Collaboration} Factor to avoid confusion.}
\begin{tabular}{lcc}
\hline
Factor & Av.g. Score (SD) & Avg. Factor Count \\
\hline
Exploration & \textbf{18.21} (1.76) & \textbf{4} \\
Enjoyment & \underline{17.85} (1.81) & 1.79 \\
Expressiveness & 16.71 (2.58) & 3.07 \\
Results Worth Effort & 16.57 (2.08) & \underline{3.57} \\
Immersion & 15.07 (3.01) & 2.07 \\
\hline
Overall CSI Score & 82.16 (5.14) &  \\
\hline
\label{tab:csi}
\end{tabular}
\end{table}

Participants generally found \sys{} usable (\textsc{Easy to Use}: 5 Strongly Agree, 5 Agree, 3 Neutral, 1 Disagree; \textsc{Easy to Learn}: 7 Strongly Agree, 4 Agree, 3 Neutral). Five participants who answered neutral or disagreed explained there was not enough time to learn and utilize all the features to their potential. P6 said: ``The program was fairly easy to use and intuitive. I feel it had a lot more to offer, but we had limited time.'' 

All of them except one agreed they would use the system for creative writing tasks (6 Strongly Agree, 7 Agree, 1 Neutral) and for exploring ideas in general (9 Strongly Agree, 4 Agree, 1 Neutral). They also found \sys{} provided creativity support, scoring CSI at 82.16 (SD = 5.14) --- a high score for creativity support index~\cite{cherry2014quantifying}. As shown in Table~\ref{tab:csi}, while \sys{} scored relatively high in all dimensions, participants found it provided the strongest support for the exploration of ideas (M = 18.21, SD = 1.76) and regarded it as the most important aspect (Avg. Factor Count = 4). 

\subsection{How Can \sys{} Be Used in Creative Workflows?}
We report three workflows (\textbf{Ws}) observed from the study, demonstrating the flexibility and scalability of our system and framework. 
\newline

\noindent\textit{W1. Using One or Multiple Responses from a Single Prompt (1/14).} P2 found that a single batch of responses provided enough for his creative writing task. For example, tasked with copywriting on the topic of `Latest Trend of UI Design across the Globe,' P2 selected responses to UI design trends varying from Asia to South America. He incorporated three responses covering three different continents (Asia, South America, and Oceania) into the text editor as three paragraphs, then edit and merge them. This workflow suggests that a single batch of responses can provide enough assistance for some creative writing tasks. 
\newline

\noindent\textit{W2. Using Responses Generated from a Chain of Prompts (8/14).} Most participants employed a chaining of prompts, whereby they progressively refined their creative output by iterating through prompts and responses. As P3 was writing a short story about `forgotten memories,' P3 first found a story about friendship and then asked AI to generate a detailed profile of the group of 5 friends. Eventually, the participant prompted AI to generate a final closing paragraph on the moral insight of the story. This iterative approach enabled P3 to build upon numerous ideas and refine his initial ideas, ultimately leading to complex and nuanced creative outcomes.
\newline

\noindent\textit{W3. Using Responses Generated via Dimension-Driven Generation Techniques (5/14).} Several participants leveraged novel features provided in \sys{} for generating additional responses. P4, who wrote `Adventures of Tom Sawyer, the Time Traveller', added label \textcolor{ACMPurple}{\code{future}} in the \textsf{Time Period} dimension during her exploration to generate story versions that have the \textcolor{ACMPurple}{\code{future}} element. P7 also used the same feature to generate new results. Tasked with writing an email where the instruction read, `write an email to ask a professional translator to step down from a project,' P7 steered the system towards producing responses that aligned more closely with her evolving creative vision, allowing her to get the version that matched the desired \textsf{tone} and \textsf{formality}.

\subsection{How Does \sys{} Help with Divergent Thinking?}

The analysis of participants' survey, interview responses, and exploration suggest that \sys{} helps users engage in divergent thinking. On questions `The tool was helpful in allowing me to track different ideas, outcomes, or possibilities' and `It was easy for me to explore many different ideas, options, or outcomes, using this tool,' \sys{} received an average score of 9/10 (1 as Strongly Disagree, 10 as Strongly Agree; SD = 1.1). 

\begin{figure*}
    \centering
    \begin{subfigure}{\textwidth}
        \includegraphics[trim=0cm 0cm 0cm 0cm, clip=true, width=445pt]{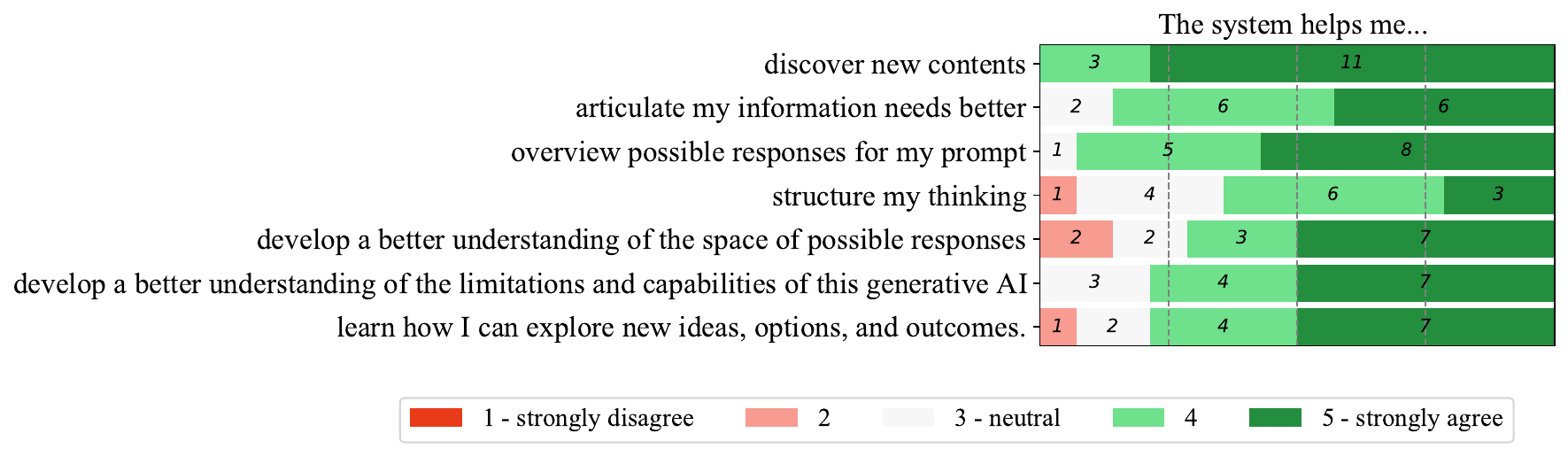}
        \caption{Evaluation of the system}
        \label{fig:system-eval}
    \end{subfigure}
    \begin{subfigure}{0.65\textwidth}
        \centering
        \includegraphics[trim=0cm 0cm 0cm 0cm, clip=true, width=340pt]{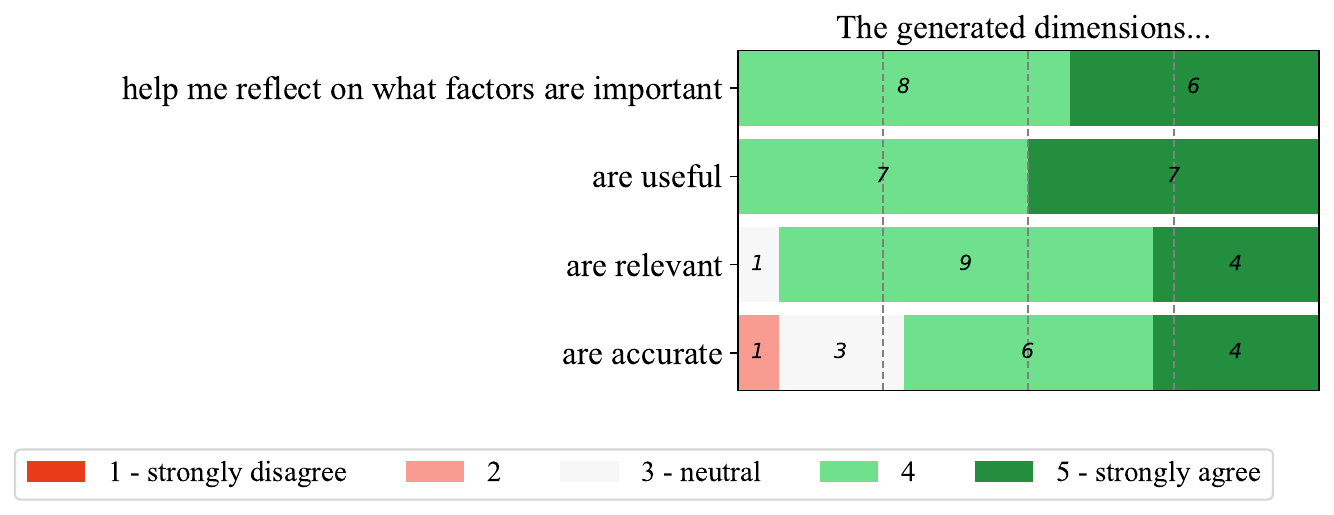}
        \caption{Evaluation of dimensions generated from the prompt}
        \label{fig:dimension-eval}
    \end{subfigure}
    \caption{Evaluation of the system and the dimensions generated from the prompt}
\end{figure*}

\textit{It helps by generating multiple responses even from a simple prompt.} All the participants reported that having multiple options from one prompt was beneficial for the brainstorming and ideation stage of creative writing. They found that it offers variation, serendipitous responses (P2), and the ability to mix-and-match different responses (P5) and think outside the box (P10). P2 said: ``This AI gives a lot of ideas and options to choose, from just one simple prompt. I think this is very powerful.'' As shown in Fig.~\ref{fig:system-eval}, all the participants also agreed that the system helped discover new content (11 Strongly Agree, 3 Agree). For example, P3 said having access to a diverse set of ideas inspired new ideas: ``while seeing these options, another ideas come.'' On average, each participant prompted 4.3 times (SD = 3.9, range = [1, 15]) and explored 13.8 responses (SD = 7.6, range = [4, 26]) during the 25-minute study session.
As a result, participants suggested that \sys{} can help avoid fixation. P6 said:

\begin{displayquote}
``It gives you a range of possible scenarios. This broadens your mind so you won't be fixed on something. I think it's the way. It gives me options and lets me know there are multiple pathways I can choose.''
\end{displayquote}

\textit{It helps with thinking outside the box.} In addition to benefiting from a sheer number of options, participants also felt the responses were creative, saying that \sys{} ``showed ideas [they] did not think of'' (P5) and ``helped think outside the box'' (P2). P2 elaborated by saying that whereas what we can see is constrained by a limited set of experiences and contexts, ``AI can go 1,000 or 1,000,000 ways different than [our] context because it has all the intelligence it learned over time.'' She further highlighted that it can broaden our horizon by making different perspectives more accessible, saying: ``it is a good way not only to enhance creativity, but also to see other points of view from different kind of people, from different kind of situations that obviously you do not have access to.''

\textit{It helps by shifting the focus from prompt engineering to exploring and generating new ideas.} The availability of options and their creativeness made it possible for participants to break free from the iterative prompting and focus instead on exploring and generating new ideas. P6 said: ``I like this better than just having one option and then me having to fine tune everything or tweak it up to my liking. This way, I have more choice and more freedom. I don't want to just get something spit out and then rely on it.''

\textit{The flexible interactions and exploration view enable creative exploration.} Many found the various interactions in \sys{} useful for creative exploration. P5 highlighted the ease with which she could vary the responses, saying: ``there's so many customization options and it generates even more ideas that than you could have thought of. You are able to generate different elements of the story like the plot, the language, the tone, the characters, the setting. And you can keep customizing and the result it brings out actually makes sense.'' In addition to flexible interactions, several mentioned the exploration view as their favorite feature as it enabled creative exploration. They elaborated, saying they enjoyed seeing ``everything'' (P1) and ``more area of thinking'' to see ``variations'' and ``expand the story'' (P3).

\subsection{How Does \sys{} Help with Developing an Understanding of the Design Space?}
\label{sec:understanding-design-space}

Participants generally agreed that \sys{} helped them develop a better understanding of the space of: (1) possible responses (7 Strongly Agree, 3 Agree, 2 Neutral, 2 Disagree); and (2) the limitations and capabilities of the AI (7 Strongly Agree, 4 Agree, 3 Neutral). Participants who disagreed or answered neutral attributed the reason not to \sys{} but to lack of time. For example, P8 said: ``These are useful, but I would need more time because I am a slow learner. I would need to spend maybe a week just playing around.'' Participants also agreed that using \sys{} ingrained in them a sense that they can now foresee possible response space of their prompts (8 Strongly Agree, 5 Agree, 1 Neutral). On average, participants selected the dimension button 1.8 times (SD = 2.1, range = [0, 7]), exploring 1.1 unique (SD = 1.2, range = [0, 4]) 1-dimensional and 1.2 unique (SD = 1.7, range = [0, 4]) 2-dimensional design spaces.

\textit{Dimensions help with reflecting on the design space.} All participants found that our approach, especially the generated dimensions, helped them reflect on what factors are important for their task and topic (6 Strongly Agree, 8 Agree). Participants mostly found the generated dimensions to be accurate (4 Strongly Agree, 6 Agree, 3 Neutral, 1 Disagree) and found them very useful (7 Strongly Agree, 7 Agree) and relevant (4 Strongly Agree, 9 Agree, 1 Neutral) for exploring ideas and reflecting on the design space. Moreover, it reminded them of aspects of the task they had never thought about. P7 said: ``it made me think about the \textsf{level of empathy}, the positive or negative aspect of [\textsf{tone}] that I never think about.''

\textit{Dimensions help with structuring thinking around the design space.} Most participants found that \sys{} helped them structure their thinking around the design space (3 Strongly Agree, 6 Agree, 4 Neutral, 1 Disagree). 
P1 and P5 explicitly attributed this to dimensions. 
P5 explained that dimensions helped break down different elements of the storyline and even in parts where she did not expect to help her think about the design space, saying:

\begin{displayquote}
``I find it fascinating that it was able to break down different elements of the storyline... the different story blocks, different elements that make up the story, like the setting, the character, the time, the tone and all of that. I was not expecting that there would be a tool that could really pull them out. You give it a prompt and it pulls out several different story elements and there are so many ideas to explore. I did not expect there would ever be a tool like that. And then it categorizes all of this into different elements and it's really impressive... breaking down into different dimensions helps me think about the story.''
\end{displayquote}

\textit{Exploration space helps understand the design space.} All participants agreed that the exploration view in \sys{} was also important in helping them achieve a comprehensive understanding of the design space (7 Strongly Agree, 7 Agree). They also agreed that interactively building new design spaces by selecting dimensions was not only useful for building a comprehensive understanding of the space (3 Strongly Agree, 10 Agree, 1 Neutral) but also for exploring the space of possible responses (7 Strongly Agree, 5 Agree, 1 Neutral, 1 Disagree). P6 praised this ability to explore in the space, saying:

\begin{displayquote}
``The brainstorming aspect of it is very, very helpful because it shows you all bunch of options right there. It helps you navigate exactly where you want to go with a certain topic and that's what I would predominantly use it for.''
\end{displayquote}

\subsection{How Does \sys{} Compare to Existing Approaches? What Benefits or Challenges Does This Approach Introduce?}
At the end of our interview, we asked participants to share: (1) how their workflow with \sys{} differed from their workflow with other AI tools they used for creative writing, (2) which they preferred, and (3) why. While our participants had exposure to various AI tools, because ChatGPT was the tool they used the most, all of them compared \sys{} with ChatGPT.

\textit{It is better for creative tasks.} Most participants (10/14) expressed preference for \sys{} for creative writing. P7 explained that he preferred ChatGPT, because it is simple and he did not need a long and complicated response for the writing tasks he engages in. Participants also suggested that the scenario where \sys{} fits may be contingent on the flexibility of the requirement for the writing task and the task complexity. If the project is more open-ended, \sys{} is better because it gives writers the opportunity to aberrate from their normal path. For example, P1 noted that ``for projects that are creative and flexible, [such as] brainstorming or working in marketing campaigns, art development, it could be amazing.'' P14 was particularly impressed with the way ideas are organized for creative exploration, saying:

\begin{displayquote}
    ``Where it blew every other tool out of the water is being able to present a range of options. Not just present them, but being able to present them in a way where you're not just sifting through pages of text. You can see everything and move around by categories. As a brainstorming tool, it's crazy.''
\end{displayquote}

The way \sys{} generates many diverse responses was also seen as an approach that better aligns with how they think and work as creative professionals. P8 said: ``The AI first determines some important aspects and then give you 40 different responses; iterations like this is a technique I use a lot as an artist.'' P3 testified this is also the case for writing: ``as a writer, I always look for more and more options, so it is good that I can explore all the options and then select one.''

\textit{{It simplifies the workflow by providing dimensions as an outline.}} Several participants shared they generally write an outline first and then use AI to expand on it. They said \sys{} simplifies this workflow, equating \sys{} generating dimensions as creating an outline. P5 said: ``in my current workflow, I create an outline first. A structure if I don't already have that provided to me. Then I use AI to generate the content based on what I already have. But with a tool like this, I feel like it would save you the time of creating an outline.'' 
P4 envisioned that dimensions would make her workflow more efficient and improve her writing, saying: ``I'll take advantage of the dimensions to simulate the best way to go about my write up. It would make it easier to finish on time.''

\textit{It empowers users as it expands their prompt and allows them to steer the response generation.} Participants also appreciated the flexibility and steerability of \sys{}. P5 found that she was able to customize the generation to produce a desired writing, saying: ``I was able to explore more ideas and customize storyline. What was generated at the end matched everything that I had selected --- every prompt and every idea.'' The way \sys{} expands a prompt was also seen as a unique feature that encourages deeper thinking. P10 said:

\begin{displayquote}
``ChatGPT just brings out one story. It doesn't care if you want to talk about different time periods like future,  modern time, or ancient time. It can be very limiting having to read a response and say, `I want to edit one or two things here.' You can't think too deeply this way. With \sys{}, however, you get to explore different options, so it'll make you think better.''
\end{displayquote}

\textit{There is still room for improvements such as minimizing cognitive overload.}
Although many appreciated being able to see multiple responses from a prompt and felt the agency of choosing a response from multiple alternatives outperforms having only a single response --- especially in the brainstorming stage, several found 40 responses overwhelming. P4 and P8 suggested showing fewer and simpler responses at first and offering more suggestions once users have a better understanding of their needs. P4 noted that there can be two modes: basic and advanced, where users receive fewer suggestions in the basic mode but more suggestions and access to the features in \sys{} in the advanced mode. P4 also mentioned a recommendation idea, stating: ``Perhaps the AI can be structured to suggests options that are the best fit for the story, it will help separate the wheat from the chaff and assist users in structuring their thinking.''

\subsection{Prolonged Use in the Wild: Do Participants Still Find \sys{} Useful After Using It in the Wild for an Extended Period of Time?}
\label{sec:prolonged-use}

After the study, participants showed strong interest in \sys{}, asking when the tool would become publicly available. Simultaneously, we had multiple participants who said there was too little time to assess \sys{}. We saw this as an opportunity to address, to some extent, the limitation of our study where our evaluation is based on participants' limited exposure (25-50 minutes) to \sys{} and under the controlled setting. Thus, we contacted the participants, asking whether they are interested in using it freely for the duration of about 2 weeks and share their experience.

Eight participants, which included those who expressed interest in using the tool and those who said there was little time to assess it, accepted our invitation. One of the authors first met with each participant to remind them about the features and give them a URL link to \sys{}. Five participants met with this author twice and the other three met once over the duration of two weeks to share their experience. Participants received \$20 for attending each interview session, which took, on average, 28 minutes. Below, we share a summary of the insights from this in-the-wild study.

\textit{Participants still prefer \sys{} for brainstorming and creative tasks even after prolonged use.} While some preferred ChatGPT for tasks that need longer, detailed responses, all the participants still maintained that they prefer \sys{} for brainstorming and creative tasks. For example, after having used it for two weeks, P4 said, ``\sys{} right now is actually my go to for ideas when crafting stories.'' P8, who called himself ``a slow learner'' and needs ``maybe a week'' to play around with it to understand its usefulness, noted after a week of use in the wild that he finds it useful, saying: ``Now I know how to use it a little better. I think it is a very useful tool. I'm now interested to find out what I can do with it.'' Additionally, P14, a writer with 30+ years of creative writing experience, maintained the same enthusiasm he had for \sys{} from the user study to after the in-the-wild study. He was given access to \sys{} for more than two weeks, during which he extensively experimented with its limits and potential use cases. After the in-the-wild study, he was able to pinpoint the value of \sys{} in supporting a specific stage in his workflow, saying:

\begin{displayquote}
    ``I think ChatGPT is useful at the mid level of designing something. It's not good at polishing, it's not creating an end product, but when you already have your idea and you want to build a little bit of a framework for it, that's what's really useful for it. [\sys{}] seems useful for before that, where you're starting from a blank page and you just want a pool of ideas to start piecing together. It seems more like an interesting brainstorming tool that I would use during pre-production rather than during production.''
\end{displayquote}

He further expressed a desire to continue using the tool and learning about other ways \sys{} can be used, saying:

\begin{displayquote}
``I would particularly be interested in the upcoming beta tests because I'm really curious to see how other people are using it. I feel like I know the basics now. This is the point where I'd traditionally start looking to forum/Discord discussions in search of advanced applications for the tool that I hadn't thought of yet.''
\end{displayquote}

\section{Discussion}

In addition to contributing a framework for systematically generating and exploring the design space constructed by LLMs, our work demonstrates how humans and LLMs can complement each other during creative exploration. As generative models, LLMs excel at quickly generating information. Humans, on the other hand, are slower in producing a variety of responses (that vary in tone, composition, etc.) but excel at comprehending a broader context and the requirements of the tasks at hand. They can judge which response is best suited for a given task. Additionally, the black-box nature of LLMs presents a major challenge for users when it comes to instructing the model to produce the desired output. In our approach, users gain an enhanced ability to navigate the design space and evaluate AI-generated responses using dimensions. These dimensions --- along with novel features for seamlessly adding user-defined dimensions and their values to generate additional responses (and update existing responses) --- provide a structured framework for users to flexibly express their preferences and guide AI towards generating responses that are not only creative but also well-suited to their task and creative goals.

\subsection{Limitations}
We describe several limitations in the study, to clearly define the scope of our study findings and motivate future work.

\textit{Lack of control over the length and number of responses}.
As explained in Section~\ref{sec:procedure}, we intentionally fixed the number of words in each response to 150 words and number of responses to 40 to support wide exploration in their writing. But we found some participants expressed a desire for longer responses, and some even wanted fewer responses. This may have affected \sys{}'s perceived value in supporting writing tasks, causing participants to favor conventional AI tools such as ChatGPT for longer and more detailed outputs. Enabling users to adjust the number and length of responses could change participants to view \sys{} as more versatile and broadly adoptable than shown in our studies.

\textit{Lack of testing with diverse user groups.} Our study focused on creative writers with experience in using AI for their writing. Testing with diverse groups of participants such as non-professional writers would provide a more comprehensive understanding of the usefulness and scalability of our structured multi-output approach.

\textit{Lack of comparison with existing AI tools.} Finally, our attempt to compare \sys{} with other AI tools ended up being a comparison with ChatGPT based on participants' self-reports. A rigorous study comparing the two with specific measures would provide deeper insights to where \sys{} stands in the landscape of AI writing tools. Admittedly, given the ever-increasing number of AI tools in the market~\cite{futurepedia}, a rigorous comparison with other AI tools may, in practice, be infeasible. Regardless, it is worth acknowledging that our study does not provide an answer to this question.

\subsection{Future Work}

\subsubsection{Addressing Study Limitations} As noted above, there are several limitations that should be addressed through: a study with a more adjustable interface, tests with diverse user groups, and comparing our approach, to the extent possible, with alternative approaches found in other AI tools.

\subsubsection{Exploring New Interaction Paradigms Beyond Prompt Engineering} In light of recent research, which has predominantly focused on exploring ways to support and enhance prompt engineering, our work raises an interesting question: ``Can we extend our interactions with LLMs beyond prompt engineering?'' While our \textsf{Prompting for Design Space} framework was designed to enhance interactions with LLMs in creative processes, we believe this work can serve as a source of inspiration, stimulating new lines of research aimed at envisioning novel interaction paradigms across various workflows and contexts. Although LLMs have enabled users to execute complex actions through natural language, prompt engineering does not always provide intuitive or efficient means to perform complex actions~\cite{zamfirescu2023johnny}. This work shows that there could be a rich opportunity for future research in this direction.

\subsubsection{Extending the Framework and Interaction Techniques to Other Domains and Tasks}
In this work, we have demonstrated a framework for supporting structured generation and exploration of design space in the domain of writing support. However, we believe the proposed framework can be broadly applied to a wide range of domains and tasks, as the idea of generating multiple, diverse ideas that serves as motivation of this work can generalize~\cite{dow2010parallel}. For instance, our framework could go beyond text input and output to other types of media such as image and video as input and output. For example, we could have a system that takes image as an input, generates dimensions relevant to the subject or setting in the image, and use their attributes to generate diverse images for users to explore. 
At the same time, the interaction techniques used in the framework could also be applicable (and necessary) in other domains and tasks. For example, irrespective of the output format, semantic zoom may be essential for efficient exploration and analysis of generated responses. 

\subsubsection{Implementing Design and Features to Address Cognitive Overload}
As mentioned, some participants found the multitude of outputs overwhelming, and suggested interaction techniques, such as initially revealing a limited number of outputs (e.g., 1-3 responses) and then allowing users to easily generate more as they move around the space. They also suggested offering basic and advanced modes, with the basic mode providing fewer outputs and the advanced mode offering more outputs, along with the ability to steer responses and explore the design space, as seen in the current version of \sys{}. Another suggestion for improving the navigation and organization was to allow users to dismiss or mark the responses they no longer needed. Taking these into account, we envision that the re-design of \sys{} would involve enabling flexible transition between simpler and more complex interfaces, which is in line with prior work where they also concluded the need to support flexible transition between a simpler, familiar interface and a more complex interface to accommodate various user needs and tasks that vary in complexity during different stages of the workflow~\cite{suh2023sensecape}.

\subsubsection{Generating Meaningful Dimensions}
\sys{} currently generates a set of categorical dimensions --- ordinal and nominal dimensions. While they are useful in most cases, our own testing and user study revealed that these dimensions can vary in usefulness.
For instance, \textsf{creativity} always appeared as an ordinal dimension with labels [\code{least}, \code{less}, \code{neutral}, \code{more}, \code{most}]. Yet, creative writers rarely, if at all, need or want a \code{less creative} artifact. 
Similarly, nominal dimensions occasionally demonstrated instances suggesting the need for calibration.
In response to P1's prompt \raisebox{-2pt}{\includegraphics[scale=0.125]{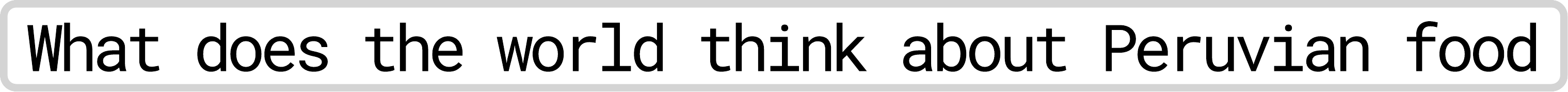}}, \sys{} generated [\code{gender}] as one of nominal dimensions. Though it was \textit{not} irrelevant to the prompt, the user did not find it useful for exploring the design space. 
While finding the set of dimensions that can satisfy everyone in every task may be impossible due to subtle variations in how one interprets the relevance of generated dimensions, the issue still affects the user experience and therefore should be tackled. A simple, quick solution would be, e.g., to provide users with the ability to easily delete irrelevant dimension(s) and use the option panel where they can specify dimension(s) to filter or check which dimension(s) were filtered out.
\section{Conclusion}

Our work addresses the challenge of harnessing the creative potential of large language models for creative endeavors. While LLMs offer vast generative capabilities, we argue that we are not leveraging them to their fullest potential by not systematically structuring the generation and exploration of their outputs. To tackle this challenge, we developed a framework --- Prompting for Design Space --- that enables the structured creation of a design space. Within this space, users can effortlessly explore, assess, and integrate numerous responses. Through the development of Luminate and a user study involving 14 professional writers, we have demonstrated the practicality and effectiveness of our framework. To our knowledge, this work is the first to explore ways to generate multiple responses from a single prompt in a systematic manner in a text-to-text context. We believe our work contributes an advancement in the way users can engage with LLMs for creative tasks, providing novel ways to harness the creative power of LLMs through structured output generation and exploration during the creative process.

\bibliographystyle{ACM-Reference-Format}
\bibliography{references}

\appendix
\newpage

\section{Appendix}
\label{sec:appendix}

\subsection{Examples of Study Participant's Creative Writing: Copywriting, Story,  Email, and Lyrics}
\textit{* Sentences or phrases with \ul{underline} indicate AI-generated content. }
\newline
\textit{** Different fonts are used to specify each prompt study participants used and its corresponding responses that were generated and used in their final text.}\newline

\noindent\textbf{Copywriting Task (P1)}
\newline
\noindent\textit{P1's Prompts:}
\begin{itemize}[label={}]
    \item \textbf{1.} \texttt{Peruvian cuisine as a cultural and social activity}
    
    \item \textbf{2.} \textit{Peruvian cuisine}
    
    \item \textbf{3.} \textsc{Proofread the text and add ideas of slangs in Spanish}
    
    \item \textbf{4.} \textbf{Which Peruvian dishes have an Asian origin}
\end{itemize}

\begin{figure}[htb!]
    \centering
    \includegraphics[trim=0cm 0cm 0cm 0cm, clip=true, width=1\columnwidth]{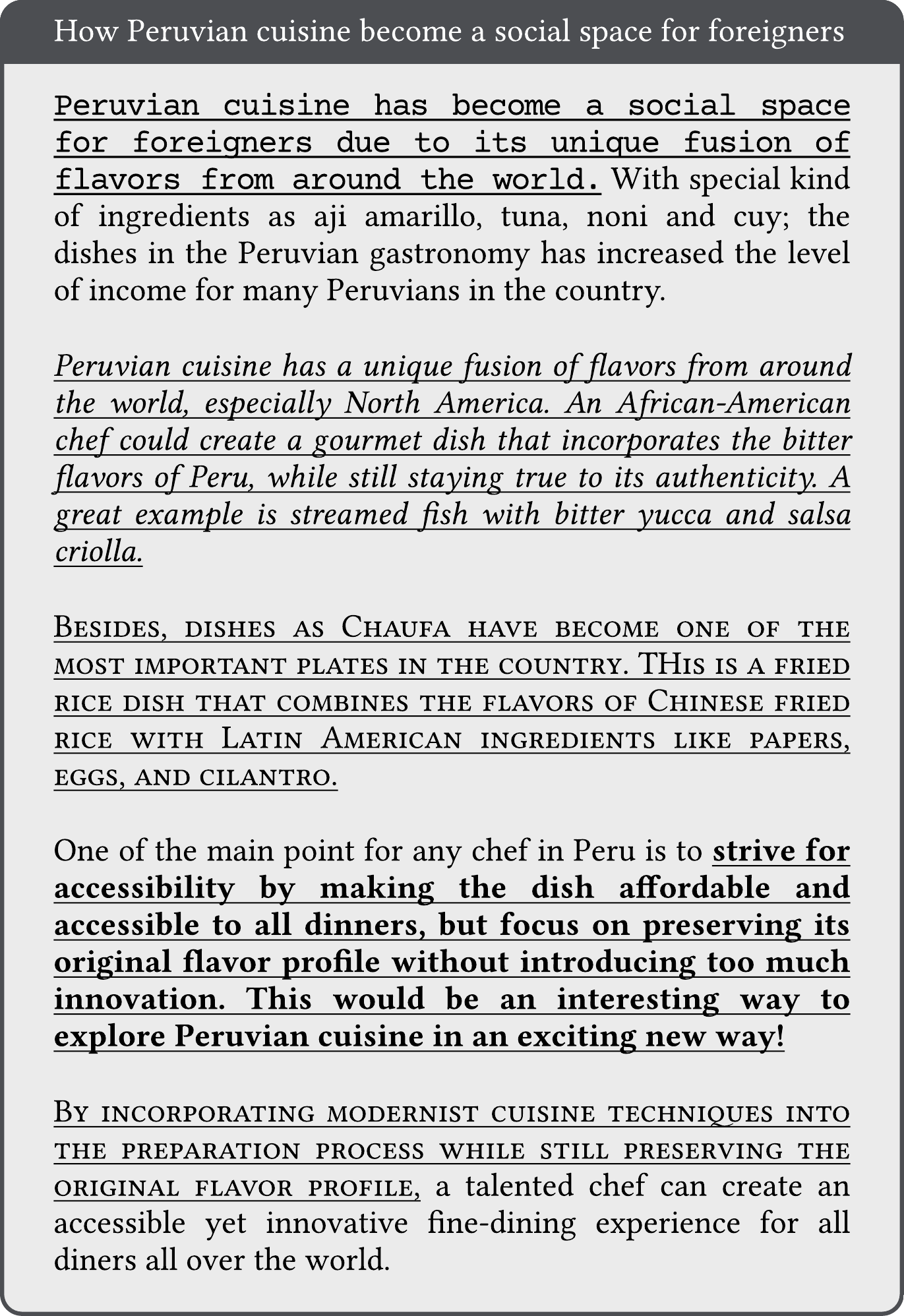}
    \label{fig:copy-writing-sample}
\caption*{ \ \ \ \ \  }\vspace*{-20pt}
\end{figure}

\noindent\textbf{Story Writing Task (P5)}
\newline
\noindent\textit{P5's Prompts:}
\begin{itemize}[label={}]
    \item \textbf{1.} \texttt{Write me a story about a man from the future specifically from the year 2079. Make it fun, thrilling, and adventurous.}
    
    \item \textbf{2.} \textit{Adding 100 more words to make the story more interesting}
\end{itemize}

\begin{figure}[htb!]
    \centering
    \includegraphics[trim=0cm 0cm 0cm 0cm, clip=true, width=1\columnwidth]{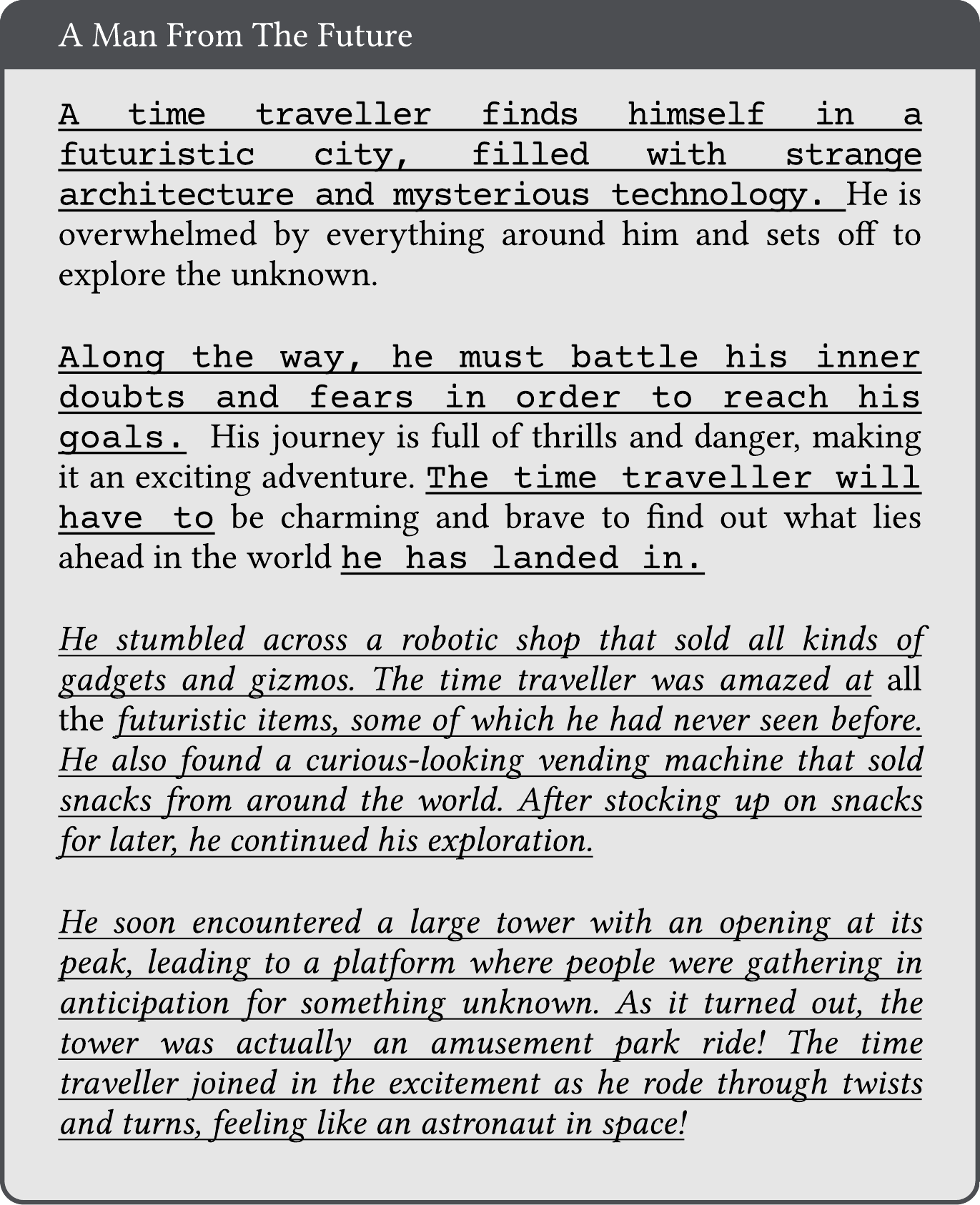}
    \label{fig:story-writing-sample}
\caption*{ \ \ \ \ \  }\vspace*{-20pt}
\end{figure}

\noindent\textbf{Email Writing Task (P7)}
\newline
\textit{* Names that appear in the following are pseudonyms.}
\newline
\noindent\textit{P7's Prompts:}
\begin{itemize}[label={}]
    \item \textbf{1.} \texttt{Can you help me create an email for a professional translator who works on a project that I am coordinating to inform him that he will not be part of the project anymore. I need the email to be very kind and don't hurt his self-esteem}
    \item \textbf{2.} \textit{I need it to be in first person. I am informing him (Tom is his name) that he will not be receiving more work for this project but he will still be considered for further projects}
\end{itemize}

\begin{figure}[htb!]
    \centering
    \includegraphics[trim=0cm 0cm 0cm 0cm, clip=true, width=1\columnwidth]{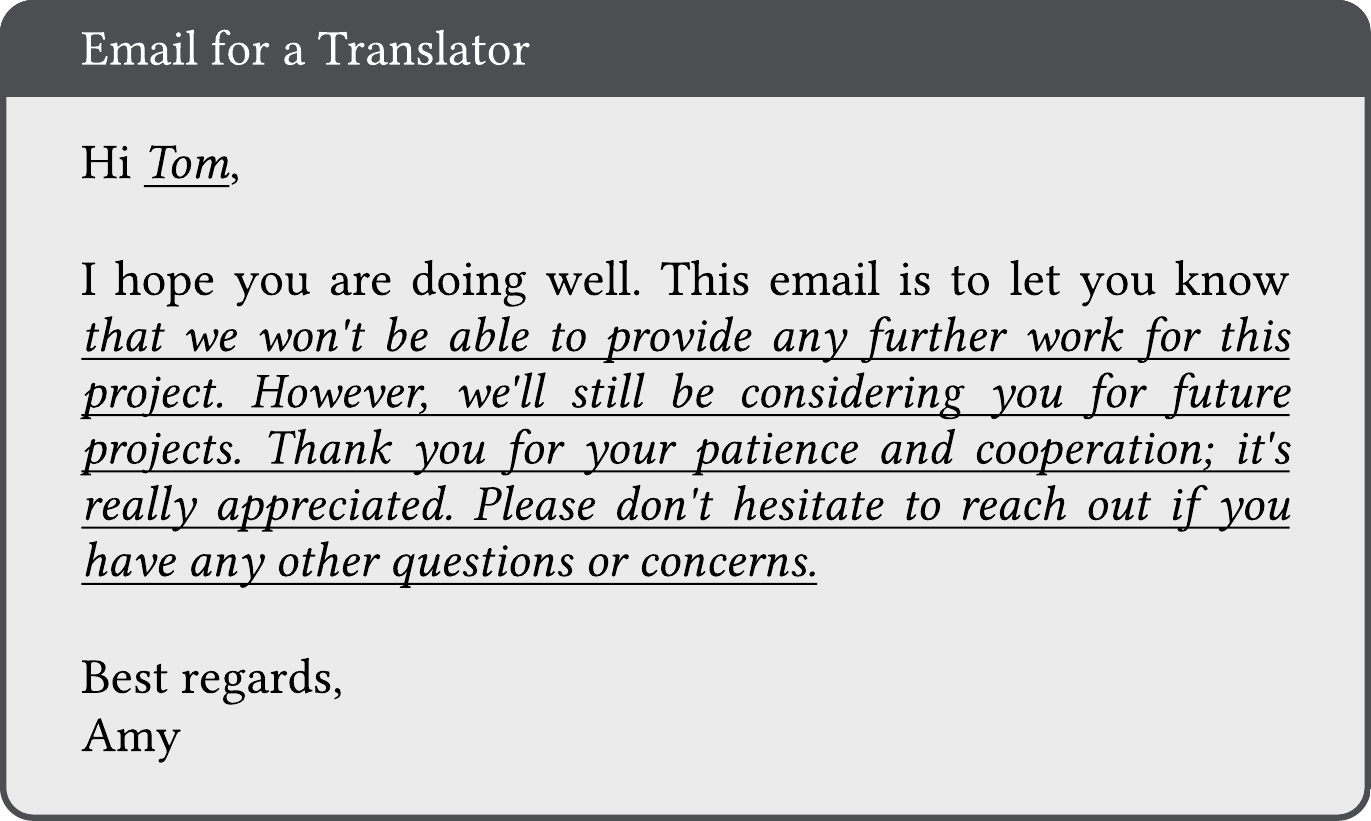}
    \label{fig:letter-writing-sample}
\caption*{ \ \ \ \ \  }\vspace*{-20pt}
\end{figure}

\noindent\textbf{Lyrics Writing Task (P8)}
\newline
\noindent\textit{P8's Prompts:}
\begin{itemize}[label={}]
    \item \textbf{1.} \texttt{Write me dark lyrics with themes of death, money, politics, industrial corporation, and ambition}
    
    \item \textbf{2.} \textit{Adding a new character to the story}
    
    \item \textbf{3.} \textsc{Add a new verse and chorus that evolves the story where this dark world and Sir William have a final conflict where the world changes but so does Sir William. The world becomes more altruistic but not too much, but Sir William loses his faith in the good in the world}
\end{itemize}

\aptLtoX[graphic=no,type=html]{\begin{figure}[htb!]
    \centering
    \includegraphics[trim=0cm 0cm 0cm 0cm, clip=true, width=0.8\columnwidth]{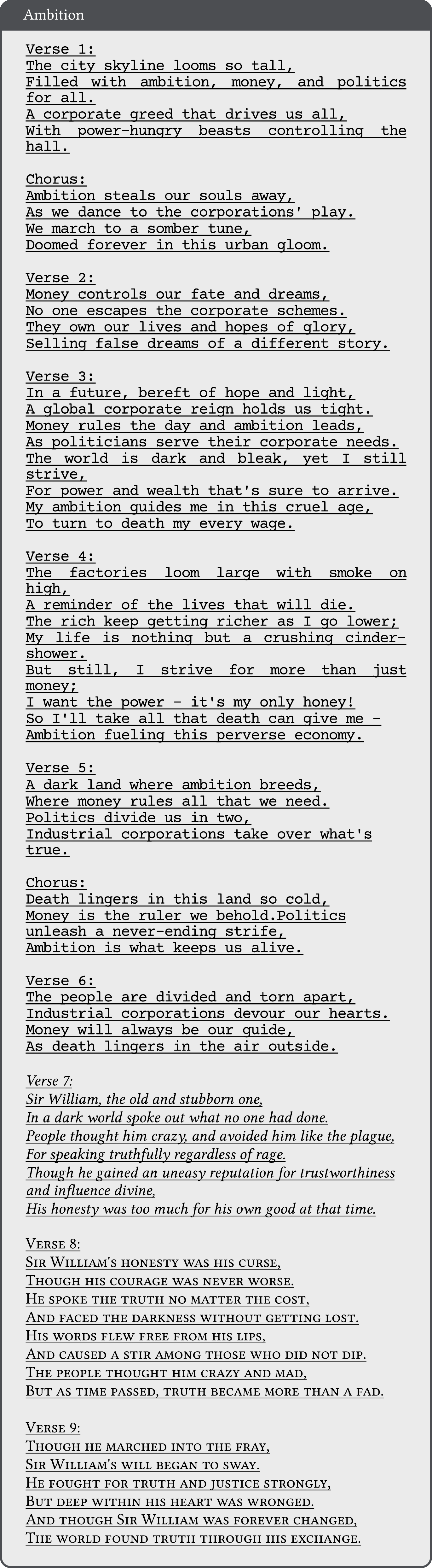}
    \label{fig:lyrics-writing-sample-2}
\caption*{ \ ,}
\end{figure}}{\begin{figure}[htb!]
    \centering
    \includegraphics[trim=0cm 0cm 0cm 0cm, clip=true, width=1\columnwidth]{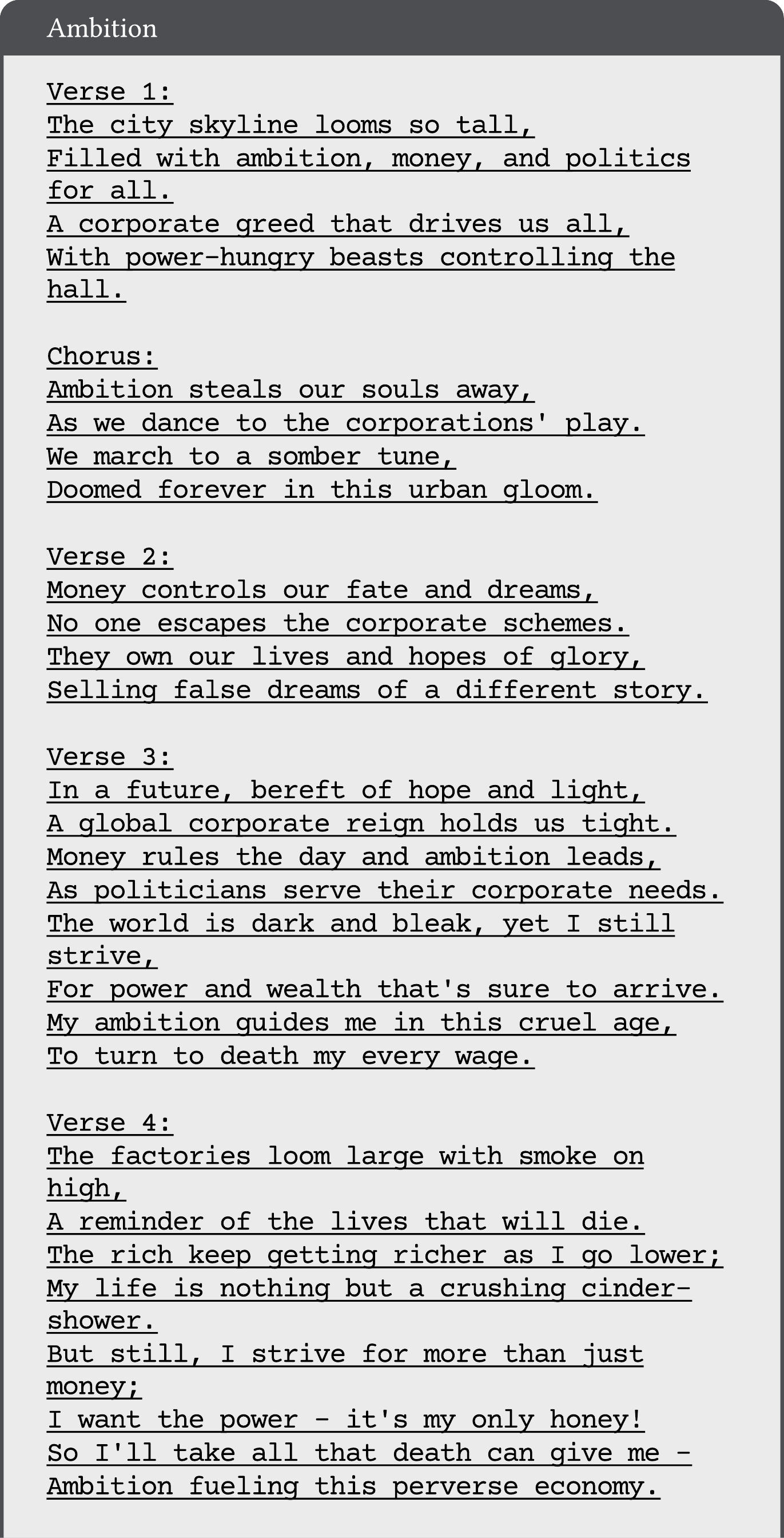}
    \label{fig:lyrics-writing-sample-1}
\caption*{ \ \ \ \ \  }\vspace*{-20pt}
\end{figure}\begin{figure}[htb!]
    \centering
    \includegraphics[trim=0cm 0cm 0cm 0cm, clip=true, width=1\columnwidth]{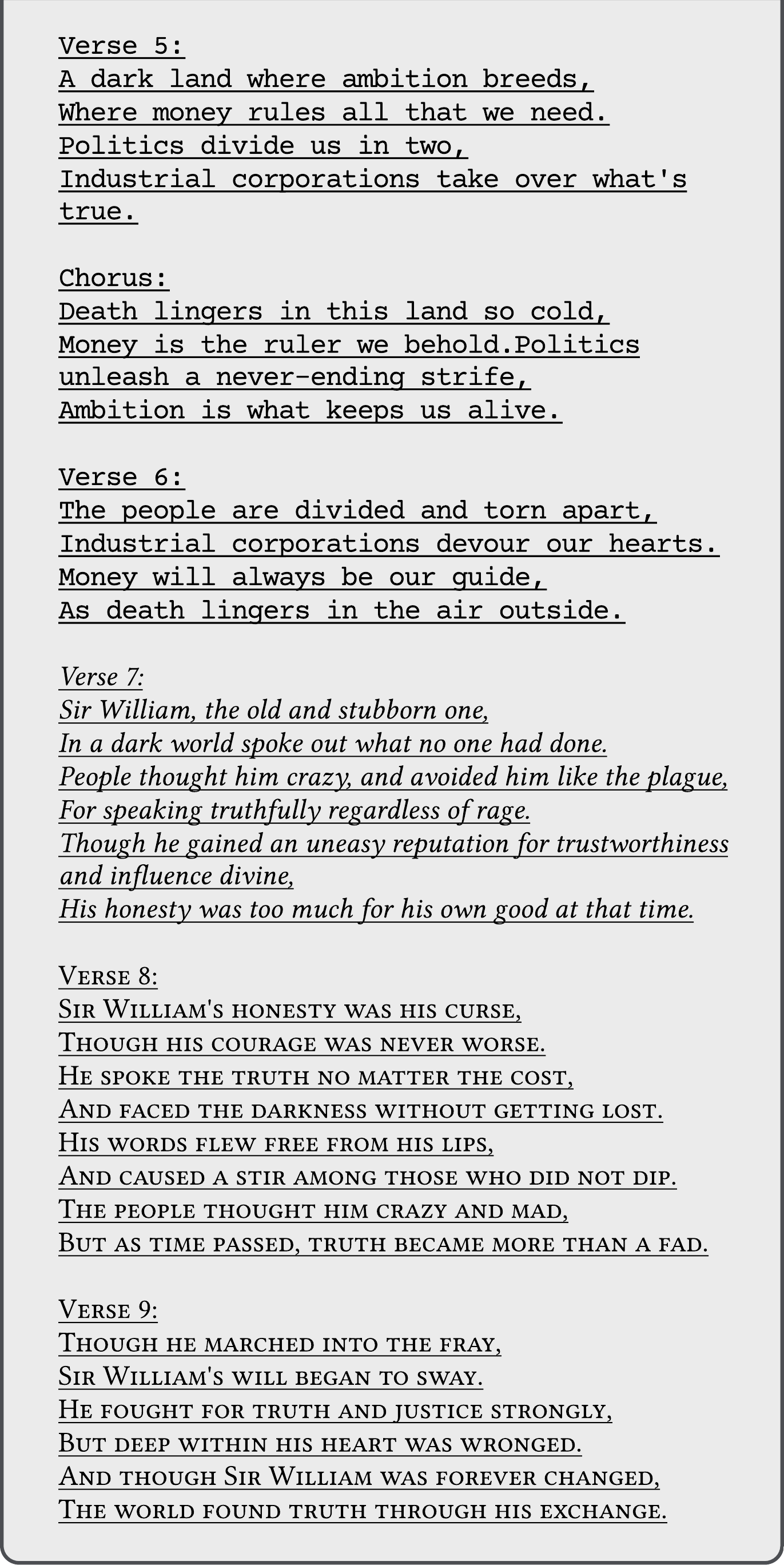}
    \label{fig:lyrics-writing-sample-2}
\caption*{ \ \ \ \ \  }\vspace*{-20pt}
\end{figure}}

\begin{table*}[htb!]
    \centering
    \caption{Time taken to generate different numbers of dimensions (Dims). The OpenAI API was invoked for different dimension quantities. The time taken to retrieve all responses was recorded over three trials (each marked as `Trial x'). Failed Call/Total Call is the ratio of failed API calls to the total number of calls. An API call is considered failed if an error occurs in the API call or the response is not in a well-formed valid JSON format (unit: seconds).}
    
    \begin{tabular}{c|cc|cc|cc|c|>{\centering\arraybackslash}p{1.4 cm}} 
        \hline
        \multirow{2}{*}{\makecell{\# of \\Dims}}& \multicolumn{2}{c|}{Trial 1} & \multicolumn{2}{c|}{Trial 2} & \multicolumn{2}{c|}{Trial 3} & \multirow{2}{*}{\makecell{Avg\\Time}} & \multirow{2}{1.4cm}{\centering\arraybackslash Failed / Total Call}\\
        & Time & Failed / Total Call& Time & Failed / Total Call & Time & Failed / Total Call & & \\
        \hline
        6& 4.91& 0/2 & 4.19& 0/2 & 4.63& 0/2 & 4.58& 0/6 \\ 
        10& 6.07& 0/2 & 8.13& 1/3 & 5.91& 0/2 & 6.70& 1/7 \\ 
        14* & 20.25& 4/6& 44.82& 6/8& 32.19& 3/5& 32.42& 13/19\\
        \hline
    \end{tabular}

        \small * Increasing the number of dimensions leads to token limits being exceeded, resulting in JSON parsing errors, such as unclosed braces.
    \label{tab:generate-dimension-time}
\end{table*}
\begin{table*}[htb!]
    \centering
   \caption{Time taken to generate different numbers of responses (Resps). The OpenAI API was invoked for different response quantities, and the time taken to retrieve all responses was recorded over three trials (each marked as `Trial x'). Failed Call/Total Call is the ratio of failed API calls to the total number of calls. An API call is considered failed if an error occurs in the API call or the response is not in a well-formed valid JSON format. For each response, the API is invoked twice (once for text generation and once for text summarization).  Thus, the total calls are equal to twice the number of responses (unit: seconds).}

    \begin{tabular}{c|cc|cc|cc|c|>{\centering\arraybackslash}p{1.4 cm}} 
        \hline
        \multirow{2}{*}{\makecell{\# of\\Resps}} & \multicolumn{2}{c|}{Trial 1} & \multicolumn{2}{c|}{Trial 2} & \multicolumn{2}{c|}{Trial 3} & \multirow{2}{*}{\makecell{Avg\\Time}} & \multirow{2}{1.4cm}{\centering\arraybackslash Failed / Total Call}\\
        & Time & Failed / Total Call& Time & Failed / Total Call& Time & Failed / Total Call& & \\
        \hline
        20 & 7.75& 0/40& 8.01& 0/40& 6.79& 0/40& 7.52& 0/120\\ 
        40 & 7.63& 1/81& 9.12& 0/80& 7.94& 1/81& 8.23& 2/242\\ 
        60 & 9.01& 0/120& 9.51& 1/121& 8.05& 0/120& 8.88& 1/361\\ 
        80* & 8.72& 9/169& 9.42& 9/169& 11.11& 9/169& 9.75& 27/507\\
        \hline
    \end{tabular}
    \label{tab:generate-response-time}

        \small * Appear ``too many request'' error when calling OpenAI API

\end{table*}

\begin{table*}[htb!]
    \caption{A list of constants defined to express the prompts used in \sys{} and shown in Table~\ref{table:prompt-list}.}
    \label{table:prompt-constant}
    \centering
\begin{tabular}{p{4.5cm} p{7.5cm}}
    \toprule
 Prompt Constant& Value\\
     \midrule
    dimensionDef& A dimension will contain categorical dimension values (attributes) that are qualitative and subjective to the user. This means there is no right answer for selecting a dimension value. The user should be able to select any dimension value depending on their preference. The dimensions must not be an evaluation of how good the writing is. All responses are assumed to be the best writing generated by you.\\
    \midrule
    dimensionConclusion& 
    Even though I encourage you to use some of these examples if best fitting, I highly recommend that I also get unique and orthogonal dimensions.\\
    \midrule
    nominalDimensionDef& 
    \textbf{dimensionDef} + A nominal dimension will contain dimension values that do not have a particular order and are up to the user's selection. Some nominal dimensions that I would want are Tone, Setting, Style, or Perspective. I do NOT want Length, Grammar, Quality, or Clarity. + \textbf{dimensionConclusion}\\
    \midrule
    ordinalDimensionDef &
    \textbf{dimensionDef} + An ordinal dimension will contain dimension values measured in an order (least, less, neutral, more, most). The type of dimensions I want are ones that are of a single key property that the user may want more of or less of depending on their preference. Some ordinal dimensions that I would want are Concreteness, Realism, or Subjectivity. I do NOT want Quality, Creativity, or Length. + \textbf{dimensionConclusion}\\
    \midrule
    wordLimit &
    Limit the response to 150 words\\
    \bottomrule
    \end{tabular}
\end{table*}

\clearpage
\onecolumn

\aptLtoX[graphic=no,type=html]{\begin{table*}
\caption{A list of prompts used in \sys{}. The \ul{underscored text} in the `Prompt' column is a placeholder for example input(s). The \textbf{bold text} in the `Prompt' column is a placeholder for prompt constants from Table~\ref{table:prompt-constant}. A Prefix refers to strings that can be added to the prompt to provide context, but these strings are not directly sent to the API. The \textsc{small caps} in the `Prompt' column is a placeholder for prompt prefix.} \label{table:prompt-list}
\begin{tabular}{p{2.5cm} p{4.5cm} p{4.5cm} p{5cm}}
 \toprule
    Prompt Type & Prompt& Example Input(s)& Example Response\\
    \midrule
    \textsc{Previous Context Prefix}& 
    \makecell[{{p{4.5cm}}}]{This is the context: \ul{background}\\---end context ---}& \makecell[{{p{4.5cm}}}]{background = It's full of surprises, that can make us smile or frown. But it always has something to teach us when we look around.} & \makecell[{{p{5cm}}}]{N/A}\\
    \midrule
    \textsc{Existing Dimension Prefix}& 
    \makecell[{{p{4.5cm}}}]{These are the current existing dimensions and their values: \ul{currentDimensions}\\}& \makecell[{{p{4.5cm}}}]{currentDimensions = ``Setting(Nominal):[Campuse Stadium, Football Locker Room, Victory Parade]\\Engagingness(Ordinal):[least, less, neutral, more, most]''} & \makecell[{{p{5cm}}}]{N/A}\\
    \midrule
    \textsc{Nominal Dimension Generation}& \makecell[{{p{4.5cm}}}]{\textbf{nominalDimensionDef} + list \ul{catNum} nominal dimensions and associated \ul{valNum} possible values
         on which we can categorize and assess the content for the prompt: \ul{prompt}\\
        \#\#\#\#\\
        You MUST answer in the following JSON object format, wrapped in curly braces. Replace all strings with <...>. There must be \ul{catNum} items in the JSON object:\\
        \{\\
        ``<dimension name \#1>'':[<\ul{valNum} values for this dimension>],\\
        ...,\\
        ``<dimension name \#\ul{catNum}>'' : [<\ul{valNum} values for this dimension>]\\
        \}\\
        }
    &
    \makecell[{{p{4.5cm}}}]{catNum = 5\\ valNum = 6\\ prompt = ``write a story about a rabbit''}
    &
    \makecell[{{p{5cm}}}]{\{\\
    ``Genre'': [``Fantasy'', ``Adventure'', ``Romance'', ``Mystery'', ``Comedy'', ``Drama''],\\
    ``Tone'': [``Lighthearted'', ``Humorous'', ``Moody'', ``Frightening'', ``Hopeful'', ``Suspenseful''],\\
    ``Setting'': [``Modern Day'', ``Medieval Times'', ``Western Era'', ``Futuristic World'', ``Mythical Realm'', ``Urban City''],\\
    ``Style'': [``Narrative Poem'', ``Dialogue Driven Story'', 
    ``Traditional Fable Tale'', ``Nonlinear Prose Piece'', ``Epic Saga'', ``Short Story''],\\
    ``Perspective'': [``First-Person POV (Protagonist)'', 
    ``Third-Person Limited (Protagonist)'', ``Third-Person Omniscient (Narrator)'', 
    ``Second-Person POV (Reader, Audience)'', ``Multiple Perspectives/Voices'', ``Objective/Impersonal Narrator'']\\
    \}}\\
    \midrule
    \textsc{Ordinal Dimension Generation}& 
    \makecell[{{p{4.5cm}}}]{\textbf{ordinalDimensionDef} + list \ul{catNum} ordinal dimensions
       on which we can assess the outcome for the prompt: \ul{prompt} to what extent represents the dimensions\\
      \#\#\#\#\\
      answer in the following JSON format: \\
      \{\\
         ``<dimension name>'': [``least'', ``less'', ``neutral'', ``more'', ``most'']\\
      \} \\
    }
    &\makecell[{{p{4.5cm}}}]{catNum = 5\\ prompt = ``write a story about a rabbit''}
    &\makecell[{{p{5cm}}}]{\{\\
  ``Creativity'': [``least'', ``less'', ``neutral'', ``more'', ``most''],\\
  ``Imagination'': [``least'', ``less'', ``neutral'', ``more'', ``most''],\\
  ``Grammatical Accuracy'': [``least'', ``less'', ``neutral'', ``more'', ``most''],\\
  ``Originality'': [``least'', ``less'', ``neutral'', ``more'', ``most''],\\
  ``Presentation Style'': [``least'', ``less'', ``neutral'', ``more'', ``most'']\\
  \}}\\
    \midrule
    \textsc{Response Generation} 
    & \makecell[{{p{4.5cm}}}]{\textbf{wordLimit}\\ 
    \#\#\#\#\\
    \textsc{Previous Context Prefix}\\
    \#\#\#\#\\
    Prompt: \ul{prompt}\\ 
    \#\#\#\#\\
    Requirements: \ul{requirements}}
    &
    \makecell[{{p{4.5cm}}}]{Prompt = ``write a story about a rabbit''\\
    Requirements = ``Genre: Comedy\\
    Tone: Frightening\\
    Setting: Medieval Times\\
    Style: Short Story\\
    Perspective: Third-Person Omniscient (Narrator)\\
    Creativity: more\\
    Imagination: most\\
    Grammatical Accuracy: less\\
    Originality: less\\
    Presentation Style: less''\\}
    &
    \makecell[{{p{5cm}}}]{Once upon a time, in a far away kingdom, there lived a brave rabbit named Flopsy. Flopsy was determined to prove himself as the bravest rabbit in the land. One day, he decided to venture into the dark and mysterious forest that surrounded his home. As he ventured deeper into the forest, he encountered all sorts of frightening creatures and obstacles. He eventually came across an old castle with an eerie presence. He cautiously entered and soon found himself face-to-face with a giant dragon! The dragon roared and Flopsy trembled in fear but he stood his ground and bravely faced off against the beast. After a long battle, Flopsy emerged victorious! He had saved his kingdom from certain destruction and was hailed as a hero by all who heard of his courageous deed.}\\
    \midrule
    \textsc{New Dimension Generation}
    & \makecell[{{p{4.5cm}}}]{\textsc{Previous Context Prefix} \\
    \#\#\#\# \\
    prompt: \ul{prompt} + \textsc{Existing Dimension Prefix} + Generate a new dimension that is new and orthogonal to the existing ones above.}
    &\makecell[{{p{4.5cm}}}]{prompt = ``write a story about a rabbit''}
    &\makecell[{{p{5cm}}}]{Tone\\}\\
    \midrule
    \textsc{Summarization}  & 
    \makecell[{{p{4.5cm}}}]{Given following text, return key words and a one sentence summary, a structure , and a title of the text.\\
      \#\#\#\#\\
      Text is: \ul{text}\\
      \#\#\#\#\\
      Don't include any text other than the json;
      Word limit of the summary text is 20 words;
      Word limit of the title is 5 words;
      Maximum 5 key words\\
      \#\#\#\#\\
      Should be in the following JSON format: \\
      \{\\
          ``Key Words'': [``<key word 1>'', ``<key word 2>'', ...], \\
          ``Summary'': ``<summary>'',\\
          ``Structure'': ``<part 1>-<part 2>-<part 3>...'',\\
          ``Title'': ``<title>''\\
    \}\\
    }
    &
   \makecell[{{p{4.5cm}}}]{ text = ``Once upon a time, in a futuristic world, there lived a rabbit. He was an adventurous soul who loved to explore the unknown. One day, he decided to take a journey and see what the world had to offer. As he hopped along his way, he encountered many strange and wonderful things. He met robots that could talk and fly, creatures that could swim in the air, and even plants that glowed in the dark! Despite all these wonders, nothing compared to the joy of discovering new places and meeting new people. The rabbit's journey was full of laughter and fun as he made his way through this strange yet exciting world. In the end, he returned home with stories of his travels that would be told for generations to come!''}
    &
      \makecell[{{p{5cm}}}]{
    \{\\
          ``Key Words'':  [``Brave'', ``Adventure'', ``Journey'', ``Creatures'', ``Love''], \\
            ``Summary'': ``A brave rabbit embarks on a journey to explore the world and finds true love in a magical kingdom.'', \\
            ``Structure'': ``Once upon a time-Journey-Encounter creatures-Finds true love-Becomes ruler of kingdom'', \\
            ``Title'': ``Rabbit's Journey''\\
    \}}\\
\end{tabular}
\end{table*}}{\begin{landscape}
\begin{longtable}{p{2.5cm} p{4.5cm} p{4.5cm} p{5cm}}
 \caption{A list of prompts used in \sys{}. The \ul{underscored text} in the `Prompt' column is a placeholder for example input(s). The \textbf{bold text} in the `Prompt' column is a placeholder for prompt constants from Table~\ref{table:prompt-constant}. A Prefix refers to strings that can be added to the prompt to provide context, but these strings are not directly sent to the API. The \textsc{small caps} in the `Prompt' column is a placeholder for prompt prefix.} \label{table:prompt-list}\\
 \toprule
    Prompt Type & Prompt& Example Input(s)& Example Response\\
    \midrule
    \endhead
    \multicolumn{4}{r}{\textit{Continued on the next page}} \\
    \endfoot
    \bottomrule
    \endlastfoot
    \textsc{Previous Context Prefix}& 
    \makecell[{{p{4.5cm}}}]{This is the context: \ul{background}\\---end context ---}& \makecell[{{p{4.5cm}}}]{background = It's full of surprises, that can make us smile or frown. But it always has something to teach us when we look around.} & \makecell[{{p{5cm}}}]{N/A}\\
    \midrule
    \textsc{Existing Dimension Prefix}& 
    \makecell[{{p{4.5cm}}}]{These are the current existing dimensions and their values: \ul{currentDimensions}\\}& \makecell[{{p{4.5cm}}}]{currentDimensions = ``Setting(Nominal):[Campuse Stadium, Football Locker Room, Victory Parade]\\Engagingness(Ordinal):[least, less, neutral, more, most]''} & \makecell[{{p{5cm}}}]{N/A}\\
    \midrule
    \textsc{Nominal Dimension Generation}& \makecell[{{p{4.5cm}}}]{\textbf{nominalDimensionDef} + list \ul{catNum} nominal dimensions and associated \ul{valNum} possible values
         on which we can categorize and assess the content for the prompt: \ul{prompt}\\
        \#\#\#\#\\
        You MUST answer in the following JSON object format, wrapped in curly braces. Replace all strings with <...>. There must be \ul{catNum} items in the JSON object:\\
        \{\\
        ``<dimension name \#1>'':[<\ul{valNum} values for this dimension>],\\
        ...,\\
        ``<dimension name \#\ul{catNum}>'' : [<\ul{valNum} values for this dimension>]\\
        \}\\
        }
    &
    \makecell[{{p{4.5cm}}}]{catNum = 5\\ valNum = 6\\ prompt = ``write a story about a rabbit''}
    &
    \makecell[{{p{5cm}}}]{\{\\
    ``Genre'': [``Fantasy'', ``Adventure'', ``Romance'', ``Mystery'', ``Comedy'', ``Drama''],\\
    ``Tone'': [``Lighthearted'', ``Humorous'', ``Moody'', ``Frightening'', ``Hopeful'', ``Suspenseful''],\\
    ``Setting'': [``Modern Day'', ``Medieval Times'', ``Western Era'', ``Futuristic World'', ``Mythical Realm'', ``Urban City''],\\
    ``Style'': [``Narrative Poem'', ``Dialogue Driven Story'', 
    ``Traditional Fable Tale'', ``Nonlinear Prose Piece'', ``Epic Saga'', ``Short Story''],\\
    ``Perspective'': [``First-Person POV (Protagonist)'', 
    ``Third-Person Limited (Protagonist)'', ``Third-Person Omniscient (Narrator)'', 
    ``Second-Person POV (Reader, Audience)'', ``Multiple Perspectives/Voices'', ``Objective/Impersonal Narrator'']\\
    \}}\\
    \midrule
    \textsc{Ordinal Dimension Generation}& 
    \makecell[{{p{4.5cm}}}]{\textbf{ordinalDimensionDef} + list \ul{catNum} ordinal dimensions
       on which we can assess the outcome for the prompt: \ul{prompt} to what extent represents the dimensions\\
      \#\#\#\#\\
      answer in the following JSON format: \\
      \{\\
         ``<dimension name>'': [``least'', ``less'', ``neutral'', ``more'', ``most'']\\
      \} \\
    }
    &\makecell[{{p{4.5cm}}}]{catNum = 5\\ prompt = ``write a story about a rabbit''}
    &\makecell[{{p{5cm}}}]{\{\\
  ``Creativity'': [``least'', ``less'', ``neutral'', ``more'', ``most''],\\
  ``Imagination'': [``least'', ``less'', ``neutral'', ``more'', ``most''],\\
  ``Grammatical Accuracy'': [``least'', ``less'', ``neutral'', ``more'', ``most''],\\
  ``Originality'': [``least'', ``less'', ``neutral'', ``more'', ``most''],\\
  ``Presentation Style'': [``least'', ``less'', ``neutral'', ``more'', ``most'']\\
  \}}\\
    \midrule
    \textsc{Response Generation} 
    & \makecell[{{p{4.5cm}}}]{\textbf{wordLimit}\\ 
    \#\#\#\#\\
    \textsc{Previous Context Prefix}\\
    \#\#\#\#\\
    Prompt: \ul{prompt}\\ 
    \#\#\#\#\\
    Requirements: \ul{requirements}}
    &
    \makecell[{{p{4.5cm}}}]{Prompt = ``write a story about a rabbit''\\
    Requirements = ``Genre: Comedy\\
    Tone: Frightening\\
    Setting: Medieval Times\\
    Style: Short Story\\
    Perspective: Third-Person Omniscient (Narrator)\\
    Creativity: more\\
    Imagination: most\\
    Grammatical Accuracy: less\\
    Originality: less\\
    Presentation Style: less''\\}
    &
    \makecell[{{p{5cm}}}]{Once upon a time, in a far away kingdom, there lived a brave rabbit named Flopsy. Flopsy was determined to prove himself as the bravest rabbit in the land. One day, he decided to venture into the dark and mysterious forest that surrounded his home. As he ventured deeper into the forest, he encountered all sorts of frightening creatures and obstacles. He eventually came across an old castle with an eerie presence. He cautiously entered and soon found himself face-to-face with a giant dragon! The dragon roared and Flopsy trembled in fear but he stood his ground and bravely faced off against the beast. After a long battle, Flopsy emerged victorious! He had saved his kingdom from certain destruction and was hailed as a hero by all who heard of his courageous deed.}\\
    \midrule
    \textsc{New Dimension Generation}
    & \makecell[{{p{4.5cm}}}]{\textsc{Previous Context Prefix} \\
    \#\#\#\# \\
    prompt: \ul{prompt} + \textsc{Existing Dimension Prefix} + Generate a new dimension that is new and orthogonal to the existing ones above.}
    &\makecell[{{p{4.5cm}}}]{prompt = ``write a story about a rabbit''}
    &\makecell[{{p{5cm}}}]{Tone\\}\\
    \midrule
    \textsc{Summarization}  & 
    \makecell[{{p{4.5cm}}}]{Given following text, return key words and a one sentence summary, a structure , and a title of the text.\\
      \#\#\#\#\\
      Text is: \ul{text}\\
      \#\#\#\#\\
      Don't include any text other than the json;
      Word limit of the summary text is 20 words;
      Word limit of the title is 5 words;
      Maximum 5 key words\\
      \#\#\#\#\\
      Should be in the following JSON format: \\
      \{\\
          ``Key Words'': [``<key word 1>'', ``<key word 2>'', ...], \\
          ``Summary'': ``<summary>'',\\
          ``Structure'': ``<part 1>-<part 2>-<part 3>...'',\\
          ``Title'': ``<title>''\\
    \}\\
    }
    &
   \makecell[{{p{4.5cm}}}]{ text = ``Once upon a time, in a futuristic world, there lived a rabbit. He was an adventurous soul who loved to explore the unknown. One day, he decided to take a journey and see what the world had to offer. As he hopped along his way, he encountered many strange and wonderful things. He met robots that could talk and fly, creatures that could swim in the air, and even plants that glowed in the dark! Despite all these wonders, nothing compared to the joy of discovering new places and meeting new people. The rabbit's journey was full of laughter and fun as he made his way through this strange yet exciting world. In the end, he returned home with stories of his travels that would be told for generations to come!''}
    &
      \makecell[{{p{5cm}}}]{
    \{\\
          ``Key Words'':  [``Brave'', ``Adventure'', ``Journey'', ``Creatures'', ``Love''], \\
            ``Summary'': ``A brave rabbit embarks on a journey to explore the world and finds true love in a magical kingdom.'', \\
            ``Structure'': ``Once upon a time-Journey-Encounter creatures-Finds true love-Becomes ruler of kingdom'', \\
            ``Title'': ``Rabbit's Journey''\\
    \}}\\
\end{longtable}
\end{landscape}}

\clearpage
\twocolumn

\begin{table*}[htb!]
    \caption{\textbf{Dimension Generation}. Ordinal dimensions and nominal dimensions generated by Luminate given example input(s) across different tasks including story writing, email writing, copywriting and song lyrics.}
    \label{table:dimension-prompts}
    \centering
    \begin{tabular}{p{1.8cm} | p{2.8cm} | p{2.4cm} | p{3.2cm}  p{3.2cm} }
        \toprule
\makecell{\textbf{Task Type}} & \makecell{\textbf{Example Input}}
& \makecell{\textbf{Prompt}}
& \makecell{\textbf{Example}\\ \textbf{Ordinal} \\ \textbf{Dimensions}}
& \makecell{\textbf{Example} \\ \textbf{Nominal} \\ \textbf{Dimensions}} \\\cline{1-5}
\makecell{Story\\Writing} 
& ``\ul{Write a story about time travelling}''
& \multirow{3}{2.5cm}{\textsc{Nominal Dimension Generation} and \textsc{Ordinal Dimension Generation} in \textsf{Table~\ref{table:prompt-list}}}
& \makecell{\textsf{Imagination}\\\textsf{Creativity}\\\textsf{Suspense}\\\textsf{Excitement}\\\textsf{Engagement}} 
& \makecell{\textsf{Genre}\\\textsf{Tone}\\\textsf{Setting}\\\textsf{Point of View}\\\textsf{Time Travel Method}}  \\\cline{4-5}
\makecell{Email\\Writing} 
& ``\ul{Write an email to all tennis club members about the next tournament}''
& & \makecell{\textsf{Compliance}\\\textsf{Impact}\\\textsf{Importance}\\\textsf{Relevance}
\\\textsf{Urgency}} 
& \makecell{\textsf{Age Group}\\\textsf{Format}\\\textsf{Level}\\\textsf{Location}\\\textsf{Type}}  \\\cline{4-5}
\makecell{Copywriting} 
& ``\ul{Write an article on the most exciting place to visit in Hawaii}''
& & \makecell{\textsf{Level of relaxation}\\\textsf{Level of fun}\\\textsf{Nature of activities}\\\textsf{Location amenities}\\\textsf{Quality of experience}} 
& \makecell{\textsf{Activity}\\\textsf{Location}\\\textsf{Price Range}\\\textsf{Size}\\\textsf{Time Frame}}\\\cline{4-5}
\makecell{Song Lyrics} 
& ``\ul{Write a song lyrics about the universe}''
& & \makecell{\textsf{Beauty}\\\textsf{Grandeur}\\\textsf{Magnificence}\\\textsf{Mystery}\\\textsf{Power}} 
& {\makecell{\textsf{Genre}\\\textsf{Instrumentation}\\\textsf{Length of Verse}\\\textsf{Mood}\\\textsf{Vocal Range}}} \\
    \bottomrule
\end{tabular}

\end{table*}

\begin{table*}[htb!]
    \caption{A list of some topics study participants could choose from during the user study.}
    \label{table:user-study-topic}
    \centering
    \begin{tabular}{p{4.5cm} p{7.5cm}}
    \toprule
 Writing Task & Topic \\
 \midrule
Copywriting  & \makecell[l]{
E-commerce and Retail\\
   - Product descriptions for online stores.\\
   - Promotions and discounts for specific products.\\
   - Creating a compelling shopping experience.\\
Travel and Tourism\\
   - Promoting travel destinations.\\
   - Writing travel guides.\\
   - Highlighting the features of a hotel or resort.\\
Health and Wellness\\
   - Advertising health supplements or products.\\
   - Writing about the benefits of a healthy lifestyle.\\
   - Creating content for fitness and wellness programs.\\
} \\
\midrule
Short Story  & 
\makecell[l]{Parallel Universe\\ 
Time Travel\\
Dystopian Society\\
First Contact\\
Family Secrets\\
A Mysterious Inheritance\\
Survival in the Wilderness\\
An Unlikely Hero\\
Love in Unexpected Places\\
A World Without Technology\\
The Power of Music\\
} \\
\midrule
Email / Letter  & 
\makecell[l]{Personal Letter Topics\\
   - Expressing Gratitude\\
   - Celebrations and Milestones\\
   - Updates on Life\\
   - Condolences\\
   - Friendship and Appreciation\\
Professional Letter Topics\\
   - Cover Letters\\
   - Resignation Letters\\
   - Recommendation Letters\\
   - Networking\\
Love and Relationships\\
   - Apology Letters\\
Social and Advocacy Letters
   - Letters to Elected Officials\\
   - Letters to the Editor\\
Educational Letters
   - College Application Letters\\
   - Scholarship Application Letters\\
} \\
    \bottomrule
    \end{tabular}
\end{table*}

\end{document}